\documentclass[aps,prb,twocolumn,showpacs,superscriptaddress,amsmath,amssymb,floatfix,10pt]{revtex4-1}  
\usepackage{graphicx}  
\usepackage{dcolumn}   
\usepackage{bm}        
\usepackage{amssymb}   
\usepackage{framed}
\usepackage{amsmath}
\usepackage{multirow}
\usepackage{hhline}
\usepackage{color}
\definecolor{bluegreen}{rgb}{0,0.2,0.8}
\usepackage{subfigure,amsmath,verbatim,moreverb}
\usepackage{tabularx}
\usepackage{adjustbox}
\usepackage{lipsum}
\usepackage{longtable}
\usepackage{booktabs}
\usepackage{rotating}

\usepackage{etoolbox}
\AtBeginEnvironment{align}{\setcounter{subeqn}{0}}
\newcounter{subeqn} %

\begin{document}

\title{Improving the performance of Tao-Mo non-empirical density functional with broader applicability 
in quantum chemistry and material sciences} 
\author{Subrata Jana}
\email{subrata.jana@niser.ac.in}
\affiliation{School of Physical Sciences, National Institute of Science Education
and Research, HBNI, 
Bhubaneswar 752050, India}
\author{Kedar Sharma}
\affiliation{School of Physics, Indian Institute of Science Education and
Research, Maruthamala, Vithura, 
Thiruvananthapuram 695551, India}
\author{Prasanjit Samal}
\email{psamal@niser.ac.in}
\affiliation{School of Physical Sciences, National Institute of Science Education
and Research, HBNI, Bhubaneswar 752050, India}

\date{\today}

\begin{abstract}

A revised version of the semilocal exchange-correlation functional [Phys. Rev. Lett. 117, 073001 (2016)] (TM) is 
proposed by incorporating the modifications to its correlation content obtained from the full high-density second-
order gradient expansion as proposed in the case of revised Tao-Perdew-Staroverov-Scuseria (revTPSS) [Phys. Rev. 
Lett. 103, 026403 (2009)] functional. The present construction improves the performance of TM functional over a 
wide range of quantum chemical and solid-state properties (thermochemistry and structural). More specifically, 
the cohesive energies, jellium surface exchange-correlation energies, and real metallic surface energies are 
improved by preserving the accuracy of the solid-state lattice constants and bulk moduli. The present proposition 
is not only physically motivated but also enhances the applicability of the TM functional. New physical insights 
with proper exemplification of the present modification which is presented here can further serve for more realistic
non-empirical density functional construction.

\end{abstract}

\maketitle

\section{Introduction}

Density functional theory~\cite{PhysRev.136.B864,PhysRev.140.A1133} is visualized as an extremely simplified version
of the complicated many-electron Schr\"{o}dinger equation. In this, all the quantum many-electron effects are embedded 
into an effective one-electron like potential comprising an unknown exchange-correlation (XC) density functional. As the 
exact analytic form of XC functional is not known. So, the central task of DFT is to approximate the XC energy/potential 
functional. Several approximations of the XC functionals with broad range of applications in quantum chemistry, solid-state 
physics, and material sciences are thus proposed~\cite{PhysRevB.23.5048,
PhysRevA.38.3098,PhysRevB.37.785,PhysRevB.46.6671,PhysRevLett.77.3865,
PhysRevB.72.085108,PhysRevB.73.235116,doi:10.1063/1.2912068,
PhysRevLett.100.136406,PhysRevB.79.075126,PhysRevB.84.045126,
doi:10.1021/ct200510s,PhysRevB.93.045126,doi:10.1063/1.5021597,
PhysRevLett.108.126402,PhysRevA.39.3761,doi:10.1063/1.476577,
doi:10.1063/1.2370993,PhysRevLett.82.2544,PhysRevLett.91.146401,
PhysRevLett.103.026403,PhysRevB.86.035130,doi:10.1021/ct400148r,
doi:10.1063/1.4789414,Sun685,doi:10.1021/ct300269u,PhysRevLett.115.036402,
PhysRevLett.117.073001,Wang8487,doi:10.1021/acs.jctc.8b00072,
doi:10.1002/qua.25224}. However, due to the reliable, quick and accurate output, the approximate semilocal XC functionals 
are widely used and the corresponding successes in quantum chemistry and condensed matter physics are undisputed~
\cite{B907148B,Peverati20120476,
doi:10.1063/1.1626543,doi:10.1021/ct0502763,doi:10.1021/ct100466k,
doi:10.1021/ct300868x,C7CP04913G,doi:10.1063/1.4971853,C6CP08761B,
PhysRevB.79.085104,doi:10.1063/1.4948636,PhysRevB.95.035118,
doi:10.1063/1.2835596,PhysRevB.84.035117,PhysRevB.79.155107,
doi:10.1063/1.5040786,doi:10.1063/1.5047863,PatraE9188,PhysRevLett.111.106401}.
In fact, approximations of the semilocal exchange-correlation functionals are proposed from various physical viewpoint. In 
density functional semilocal approximations, there are mainly two classes of approximations that have been widely used. The 
first one is known as non-empirical or semi-empirical density functional approximations which are practically useful for 
both the quantum chemists and solid-state physicists. However, heavily parametrized density functionals 
~\cite{doi:10.1063/1.2370993,Wang8487,doi:10.1063/1.476577} are also proposed but those functionals perform well within the 
parametrized test set and practically not so stable for the solid-state calculations. There are various ways for constructing 
the non-empirical density functionals. Some functionals are constructed from constraint satisfaction~\cite{PhysRevLett.77.3865,
PhysRevLett.82.2544,PhysRevLett.91.146401,PhysRevLett.103.026403,PhysRevB.86.035130,doi:10.1021/ct400148r,doi:10.1063/1.4789414,
Sun685,doi:10.1021/ct300269u,PhysRevLett.115.036402,PhysRevLett.117.073001,PhysRevB.93.045126} or exchange hole model~
\cite{PhysRevA.39.3761,PhysRevLett.117.073001} or satisfying both~\cite{PhysRevLett.117.073001}. Starting from the local 
density approximation (LDA)~\cite{PhysRevB.23.5048}, the higher rungs of XC density functional approximations are constructed 
by including the gradient of density and incorporating the Kohn-Sham (KS) kinetic energy dependency. The gradient dependent 
functionals are known as generalized gradient approximations (GGA)~\cite{PhysRevB.23.5048,PhysRevA.38.3098,
PhysRevB.37.785,PhysRevB.46.6671,PhysRevLett.77.3865,PhysRevB.72.085108,
PhysRevB.73.235116,doi:10.1063/1.2912068,PhysRevLett.100.136406,
PhysRevB.79.075126,PhysRevB.84.045126,doi:10.1021/ct200510s,PhysRevB.93.045126,
doi:10.1063/1.5021597,PhysRevLett.108.126402}. Whereas, those obtained by incorporating the KS kinetic energy functionals 
as an extra ingredient are recognized as the meta-generalized gradient approximations (meta-GGA)~\cite{PhysRevA.39.3761,
doi:10.1063/1.476577, doi:10.1063/1.2370993,PhysRevLett.82.2544,PhysRevLett.91.146401,PhysRevLett.103.026403,PhysRevB.86.035130,
doi:10.1021/ct400148r, doi:10.1063/1.4789414,Sun685,doi:10.1021/ct300269u,PhysRevLett.115.036402,PhysRevLett.117.073001,Wang8487,
doi:10.1021/acs.jctc.8b00072,doi:10.1002/qua.25224}. In this way, developments and subsequent attempt to construct very accurate 
approximations of XC functionals make DFT as a practically very appealing and widely used theory to extract the wealth of several 
observable quantities, such as thermochemical and kinematics of molecules, bond lengths and angles, reaction barrier heights,
dynamics of molecules, dipole moments, polarizabilities, infrared intensities, lattice constants of solids, bulk moduli, cohesive
energies, surface energies etc~\cite{B907148B,Peverati20120476,doi:10.1063/1.1626543, doi:10.1021/ct0502763,doi:10.1021/ct100466k,
doi:10.1021/ct300868x,C7CP04913G,doi:10.1063/1.4971853,C6CP08761B,PhysRevB.79.085104, doi:10.1063/1.4948636,PhysRevB.95.035118,
doi:10.1063/1.2835596,PhysRevB.84.035117,PhysRevB.79.155107,doi:10.1063/1.5040786, doi:10.1063/1.5047863,PatraE9188,
PhysRevLett.111.106401}. 

However, recent advances in functional developments and its applications indicate that the desired accuracy of both quantum 
chemistry and solid-state physics are achievable by satisfying more exact constraints by the non-empirical functionals. One 
such functional is proposed recently by Tao-Mo~\cite{PhysRevLett.117.073001}. The motivation of the present work follows from 
the alluring features of the TM functional. The TM semilocal functional is designed using the density matrix expansion (DME) 
techniques which are accurate for compact density i.e., atoms and molecules. To perform accurately in the case of solids, the 
slowly varying fourth-order gradient approximation is also included within the functional form. In the original TM functional, 
it is shown that the TM exchange performs differently for several solid-state properties with both the TPSS 
~\cite{PhysRevLett.91.146401} and TM correlation energy~\cite{PhysRevLett.117.073001} (designed by modifying the TPSS correlation). 
However, the TM correlation obeys more exact constraint than TPSS for the low-density limit~\cite{PhysRevLett.117.073001} and 
the TM exchange functional coupled with TM correlation performs more accurately than its TMTPSS counterpart for most of the
molecular and solid-state properties~\cite{doi:10.1063/1.4971853,C6CP08761B,PhysRevB.95.035118}. Interestingly, for jellium 
surface XC energy, the TPSS correlation performs better compared to TM correlation~\cite{PhysRevB.95.035118}. In this work, we 
seek a modification to the TM correlation energy prompted by the improvement achieved by the revTPSS correction
~\cite{PhysRevLett.103.026403} over TPSS correlation. It's done by implementing the revTPSS like modifications into the 
correlation energy of the TM functional. It is shown that the modified correlation coupled with TM exchange keeps all the good
features of TM functional intact. Additionally, it leads to noticable improvement of results for most of the thermochemical 
test sets, cohesive energies, jellium surface XC energies, and surface energies of the real metals by keeping accuracy of 
lattice constants and bulk moduli. This improvement clearly indicates that the change in correlation energy is necessary to 
perform equally well for both the thermochemical, bulk and surface properties of solids. 

To present all these modifications and functional performances, we organize this paper as follows. In the following section, 
we will present the underlying physical motivations and relevant modifications to the TM functional. Following it, we will 
assess the performance of the revised functional thus obtained in the context of thermochemical accuracy, solid-state lattice 
constants, bulk moduli, cohesive energies, jellium surface XC energies and surface energies of the real metals. We will conclude 
by discussing the results and future prospects of the proposed revision of the TM functional.    

\section{Theory}

As our starting point for proposed modification to the TM functional, we consider the TM exchange energy functional having the form
~\cite{PhysRevLett.117.073001}
\begin{equation}
 E_{xc}=-\int~\rho(\mathbf{r})\epsilon_{x}^{unif}(\mathbf{r})F_x^{TM}~d^3r~,  
\end{equation}
with $\epsilon_{x}^{unif}$ is the exchange energy density in the uniform electron gas approximation and $F_x$ is the TM exchange 
enhancement factor~\cite{PhysRevLett.117.073001} given by
\begin{eqnarray}\label{final}
F_x^{TM} = w F_x^{DME} + (1-w) F_x^{sc}~.
\end{eqnarray}
In this the DME based exchange enhancement factor $F_x^{DME}$ is  give by $F_x^{DME} = 1/f^2+7R/(9f^4)$, with $R=1+595
(2\lambda-1)^2p/54-[\tau-3(\lambda^2-\lambda+1/2)(\tau-\tau^{unif}-|\nabla \rho|^2/72\rho)]/\tau^{unif}$ and $F_x^{sc}$ 
is the slowly varying fourth-order gradient approximation (GE4) which is given by $F_x^{sc} = \{1+10[(10/81+50p/729)p+
146{\tilde q}^2/2025-(73{\tilde q}/405)[3\tau_W/(5\tau)](1-\tau_W/\tau)]\}^{1/10}$, where $\tilde{q} = (9/20)(\alpha - 1) 
+  2p/3$. In the original construction of TM functional, extrapolation is done between the compact density (i.e, DME) and 
slowly varying fourth-order density correction (sc) through a function  $w$. As for solids, the slowly-varying bulk valance 
region is important. Therefore, it is necessary to recover the correct fourth-order density gradient approximation of 
exchange. The function $w$ is given by,
\begin{equation}
    w=\frac{z^2+3z^3}{(1+z^3)^2}~,
\end{equation}
where $z=\frac{\tau^W}{\tau}$ is the meta-GGA ingredient. Due to different
behaviors of $z$, $w$ switches from the molecular or atomic systems to the slowly
varying solid-state system i.e., it switches from DME to the slowly varying
density limit. It is noteworthy to mention that near the bond center of the
molecules, $z\approx 0$ implies that $w\approx 0$. In the core and density tail
region, the systems become effectively one or two electron-like i.e.,
$\tau\approx\tau^W$, which implies $w\approx 1$. In bulk solids, where the
semi-classical gradient expansion of the kinetic energy density is valid, $w<1$.
Therefore, for the slowly varying density approximation, both the DME exchange
enhancement factor and slowly varying fourth order density gradient expansion
contribute.

Regarding the correlation content of the TM functional ~\cite{PhysRevLett.117.073001}, it is based on one-electron 
self-interaction free TPSS correlation. It includes the modified TPSS correlation satisfying (nearly-) exact
constraint in the low-density or strong-interaction limit of meta-GGA correlation~\cite{PhysRevLett.117.073001}. 
Also, this modification actually reduces the errors of lattice constant of bulk solids compared to its corresponding 
TPSS correlation. The modified TM correlation proposed by Tao-Mo is given as follows~\cite{PhysRevLett.117.073001}, 
\begin{eqnarray}
E_{c}^{TM}[\rho_\uparrow,\rho_\downarrow] & = & \int d^3r \rho
\epsilon_{c}^{\mathrm{revPKZB}}[1 + d \epsilon_{c}^{revPKZB}
(\tau^{W}/\tau)^3 ]~,\nonumber\\
\label{eq_ectot}
\end{eqnarray}
where
\begin{eqnarray}
\epsilon_{c}^{revPKZB} & = & \epsilon_{c}^{PBE}(\rho_\uparrow,
\rho_\downarrow,\nabla \rho_\uparrow,\nabla \rho_\downarrow)
 [1 + C^{TM}(\zeta,\xi) (\tau^{W}/\tau)^2]\nonumber\\
&-& [1 + C(\zeta,\xi)](\tau^{W}/\tau)^2
\sum_{\sigma}\frac{n_\sigma}{n} \tilde\epsilon_{c}~,\nonumber\\
\label{eq_ecden}
\end{eqnarray}
with
\begin{eqnarray}
C^{TM}(\zeta,\xi) = \frac{0.1\zeta^2 + 0.32\zeta^4}{\{1 +  \xi^2
[(1 + \zeta)^{-4/3} + (1 - \zeta)^{-4/3}]/2 \}^4}~. 
\label{eq_CCC}
\end{eqnarray}
Here, $\zeta = (\rho_\uparrow - \rho_\downarrow)/n$, and $\xi = |\nabla \zeta|/2(3\pi^2\rho)^{1/3}$. The parameter $d = 2.8$ 
Hartree$^{-1}$ is chosen to accurately predict the jellium surface correlation energy~\cite{PhysRevLett.91.146401}.

The TM functional is proved to be very accurate in predicting the energetic, and structural properties of atoms, molecules and 
solids ~\cite{doi:10.1063/1.4971853,C6CP08761B,PhysRevB.95.035118}. This is due to the exact constraint satisfaction of the 
exchange hole which is incorporated through $F_x^{DME}$. This is the first ever functional of its kind which extrapolates between 
the compact density and slowly varying fourth-order gradient expansion. It is noteworthy to mention that the prime contribution for 
the atomic and molecular exchange energy comes from the DME exchange term. However, there is still room to improve the performance 
of the TM functional. The motivation of the present paper follows from the work ~\cite{doi:10.1063/1.4971853,C6CP08761B,
PhysRevB.95.035118}, where it is shown that the TM exchange combined with the TM and TPSS correlations performs differently for 
the structural and energetic properties of solids. Nevertheless, it is also a common practice to improve the functional performance 
based on simple modifications on exchange and correlation form. Different modifications based on various physical motivations are 
done in this direction. For example, based on the Perdew-Kurth-Zupan-Blaha (PKZB)~\cite{PhysRevLett.82.2544} meta-GGA, the TPSS 
meta-GGA functional is proposed and later a simple modification on the TPSS exchange and correlation is done through revTPSS
~\cite{PhysRevLett.103.026403}. Also, we want to acknowledge several works on the recent modifications of the revTPSS and SCAN 
functional ~\cite{PhysRevB.86.035130,doi:10.1021/ct400148r,doi:10.1021/ct300269u, doi:10.1021/acs.jctc.8b00072}.

\begin{figure}
\begin{center}
\includegraphics[width=3.2in,height=1.8in,angle=0.0]{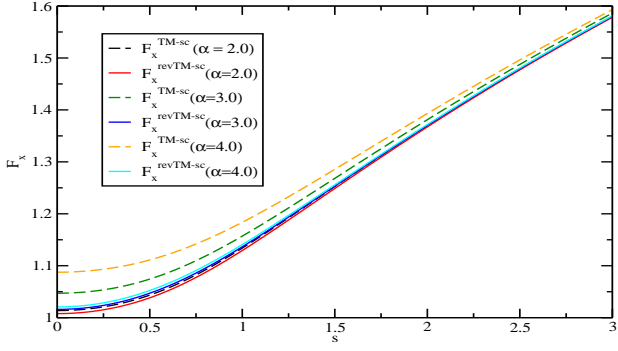}\\
\end{center}
\caption{The slowly varying enhancement factor $F_x^{sc}$ of the TM and revTM functionals with $\alpha=2.0$, $\alpha=3.0$, and 
$\alpha=4.0$.}  
\label{figaa}
\end{figure}

Now, we propose two simple modifications of the TM exchange-correlation functional which improves the molecular, bulk and surface
properties of solids keeping all the good properties of the TM functional unalter. In doing so,\\  
(i) we adopt the flexible choice for meta-GGA ingredient $\tilde{q}$ by replacing it with $\tilde{q}_b$, where $\tilde{q}_b = 
\frac{9(\alpha-1)}{20[1+b\alpha (\alpha-1)]^{1/2}}+\frac{2p}{3}$. This modification of $\tilde{q}$ is used in TPSS and revTPSS 
functional but not in TM functional. By construction, $\tilde{q}_b$ becomes $\tilde{q}$ at $b=0$. Both the $\tilde{q}_b$ and 
$\tilde{q}$ follow closely the reduced Laplacian gradient ($q$). For $0\leq\alpha\leq 1$ (single-orbital towards slowly varying 
density region~\cite{PhysRevLett.111.106401}), the $F_x^{revTM-sc}$ (exchange enhancement factor of slowly varying density 
correction of the revised TM (revTM) functional with $\tilde{q}_b$) and $F_x^{TM-sc}$ (exchange enhancement factor of slowly 
varying density correction of the TM functional with $\tilde{q}$) essentially follow the same behaviour. Whereas, for $\alpha 
>>1$ (regions of overlapping closed shells~\cite{PhysRevLett.111.106401}) and $s\approx 0$ $F_x^{TM-sc}$ and $F_x^{revTM-sc}$ 
are quite different which is relevant in the middle of bonds. On the other hand, when $\alpha$ is closer to $1$ (slowly varying 
density region), these enhancement factors are practically the same. In both cases, $F_x^{revTM-sc}$ monotonically increases 
with the reduced density gradient ($s$) (shown in Fig.~\ref{figaa}). Regarding the parameter $b$, it is chosen to be $0.40$ in 
TPSS and revTPSS based functional, so that $F_x$ becomes a monotonically increasing function of reduced density gradient ($s$).
However, in our present case, we find no reason to change this value for  modification of the TM functional. Therefore, this 
modification is done in the same spirit as is proposed from the PKZB to TPSS meta-GGA functional.   

\begin{figure}
\begin{center}
\includegraphics[width=3.2in,height=1.8in,angle=0.0]{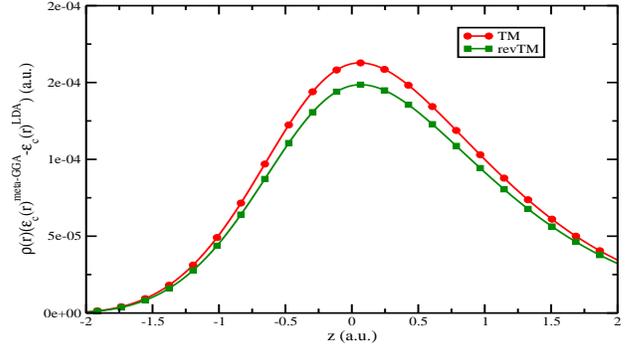} 
\end{center}
\caption{Difference of the TM and revTM correlation energy densities w.r.t. LDA}
\label{fig2}
\end{figure}

\begin{figure}
\begin{center}
\includegraphics[width=3.2in,height=1.8in,angle=0.0]{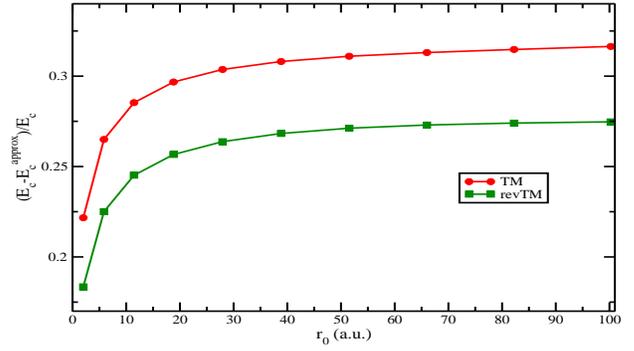} 
\end{center}
\caption{Error for the correlation energy of the Hooke's atom for the TM and revTM correlation at different values of the 
classical electron distances $r_0 = (\omega^2/2)^{-1/3}$, where $\omega$ is the frequency of the isotropic harmonic
potential. Here, $E_c$ is the exact correlation of the Hook's atom.}
\label{fig3}
\end{figure}

Having refined the TM exchange, we now focus on the TM correlation. The TM correlation is based on the TPSS correlation, 
where the parameter $\beta=0.066725$ is obtained from the high-density limit ($r_s\to 0$) of the slowly varying second-
order gradient approximation of the correlation energy functional. The TPSS correlation, which is based on the PBE 
correlation  adopts the same value of the parameter $\beta$. Later, Perdew et.al. ~\cite{PhysRevLett.103.026403} adopted 
a density-dependent $\beta$ parameter derived by Hu and Langreth~\cite{PhysRevB.33.943}, which correctly recovers the
high-density limit ($r_s\to 0$) of the slowly varying second-order gradient approximation and the low-density limit 
($r_s\to \infty$) of the second-order gradient expansion for exchange (which is $10/81$) that cancels the correlation. 
In our modification of the correlation (ii) we adopt the proposition made by Perdew et.al.~\cite{PhysRevLett.103.026403} 
in the revTPSS functional and have modified the $\beta^{TPSS}$ to $\beta^{revTPSS}(r_s)=0.066725(1+0.1r_s)/(1+0.1778r_s)$, 
where $r_s=(\frac{3}{4\pi\rho})^{1/3}$ is the Seitz radius. Apart from this modification, we keep the form of the TM 
correlation intact which is revised from TPSS correlation as~\cite{PhysRevLett.117.073001},
\begin{equation}
C(\zeta,0)=0.1\zeta^2 + 0.32\zeta^4~.
\end{equation}
This modification is actually necessary to improve the TM correlation energy functional in the low-density limit. Also, 
as the TM exchange obeys the exact fourth order gradient expansion, the choice of $\beta^{revTPSS}$ in principle can 
improve the high-density limit of the correlation through the error cancellation between second-order gradient expansion 
of exchange and correlation. Therefore, modification of the $\beta$ parameter from the high-density second-order gradient 
expansion (as is done in TPSS) to the full second-order gradient expansion (which is incorporated in revTPSS) keeps the 
conditions (i)-(iii) intact for correlation in ref.~\cite{PhysRevLett.117.073001} and additionally, in the high-density 
limit, the second-order gradient terms of the correlation cancels with exchange in the revTM functional.    

To complete our analysis, in Fig.~\ref{fig2}, we consider the correlation energy densities w.r.t. LDA as a functional 
of distance ($z$) of the jellium surface at the bulk parameter $r_s = 2$. Unlike the TM correlation, the revTM
correlation slightly de-enhanced inside the bulk solids and in the vacuum. This is logical because the revTM correlation 
uses the parameter $\beta$ which is modified from TPSS and the revTM correlation more exactly satisfies the low-density 
and high density limit~\cite{PhysRevLett.117.073001,PhysRevLett.103.026403}. Additionally, in Fig.~\ref{fig3}, we show 
the revTM correlation for the Hooke's atom. This model system is used to predict the functional performance from the
low-density or strong-interaction limit (small $\omega$) to the high-density limit (large $\omega$). From Fig.~\ref{fig3}, 
it is evident that the revTM correlation performs better than its TM counterpart due to the satisfaction of more exact 
constraint. However, more improved $\beta$ parameter for this type of model system is also proposed in TPSSloc functional
\cite{PhysRevB.86.035130} and later it is used in the BLOC functional~\cite{doi:10.1021/ct400148r}.  

\begin{table*}
\begin{center}
\caption{Summary of deviations using different methods in terms of ME and MAE. The least MAE between the TM and revTM 
are in boldfont.}
\label{thermo}
\begin{ruledtabular}
\begin{tabular}{lrrrrrrrrrrrrrrrrrrrrrr}
     &             & L(S)DA&PBE &TPSS & revTPSS & SCAN & TM &revTM  \\
\hline
      \multicolumn{9}{c}{atomic energy (Ha)} \\[0.1 cm]
AE17   & ME  &-0.673   &  -0.076     &0.028       &-0.029       &-0.220 &-0.054 &-0.054\\
       &MAE  &0.673   & 0.076      &0.028       &0.038       &0.251 &{\textbf{0.054}} &{\textbf{0.054}}\\[0.4 cm]

      \multicolumn{9}{c}{atomization energy (kcal/mol)} \\[0.1 cm]
AE6    & ME  & -76.4   &-11.8   &-3.7&-2.9&-1.1&0.1&-1.1  \\
       & MAE & -76.4   &15.0   &5.6&6.0&3.4&5.1&{\textbf{4.3}}   \\
      \multicolumn{9}{c}{electron affinity (kcal/mol)} \\[0.1 cm]
EA13   &ME   &-5.60  & -1.05 &0.76&1.42&1.01&3.20&2.39 \\
       &MAE  &5.71   &2.20  &2.37&2.69&3.30&3.72&{\textbf{3.31}}   \\[0.4 cm]
       \multicolumn{9}{c}{ionization potential (kcal/mol)} \\[0.1 cm]
IP13   &ME   & -4.48 &-2.01 &-1.69&-0.61&-3.80&0.40&-0.01  \\[0.1 cm]
       &MAE  & 5.31  &3.49  &3.02&2.91&4.47&2.99&{\textbf{2.98}}  \\[0.4 cm]
       \multicolumn{9}{c}{proton affinity (kcal/mol)} \\[0.1 cm]
PA8    &ME   & 4.99  &-0.16 &-2.81&-2.92&-1.11&-1.60&-1.52  \\
       &MAE  & 4.99  &1.42  &2.81&2.92&1.20&1.88&{\textbf{1.61}}  \\[0.4 cm]
       \multicolumn{9}{c}{barrier heights (kcal/mol)} \\[0.1 cm]
BH6        &ME   &-17.9   &-9.42  &-8.34&-7.43&-7.49&{\textbf{-7.45}}&-7.65  \\
 &MAE   &17.9   &9.42   &8.34&7.43&7.49&{\textbf{7.45}}&7.65  \\
        \multicolumn{9}{c}{hydrogen bonding (kcal/mol)} \\[0.1 cm]
HB6        &ME   & -4.50  &-0.12  &0.43 &0.40 &-0.93 &-0.13 &0.12  \\
 &MAE   & 4.50  &0.32   &0.43 &0.40 &0.93 &0.20 &{\textbf{0.18}}  \\
         \multicolumn{9}{c}{dipole bonding (kcal/mol)} \\[0.1 cm]
DI6        &ME   & -2.92  &-0.33  &0.35 &0.30 &-0.60 &-0.50 &-0.17  \\
 &MAE   & 2.92  &0.40   &0.51 &0.47 &1.12 &0.50 &{\textbf{0.37}}  \\

       
       \\
       
\end{tabular}
\end{ruledtabular}
\end{center}
\end{table*}

\section{COMPUTATIONAL DETAILS}
To assess the performance of the revTM functional with other popular GGA and meta-GGA functionals, we implement the revTM
functional in NWChem~\cite{VALIEV20101477} and Vienna Ab initio Simulation Package (VASP) ~\cite{PhysRevB.47.558,PhysRevB.54.11169,
PhysRevB.59.1758,KRESSE199615} code for the molecular and solid-state calculations respectively. All molecular and solid-state 
calculations are done self-consistently using the $6-311++$G($3$df,$3$pd) basis set in NWChem and plane-wave basis set in VASP. For 
molecular calculations we consider Minnesota $2.0$~\cite{Peverati20120476} test set which includes (i)AE$17$ $-$atomic energies of
$17$ atoms (H$-$Cl), (ii) AE$6$ $-$ atomization energies of $6$ molecules, (iii) EA$13$ - $13$ electron affinities, (iv) 
IP$21$ $-$ $21$ ionization potentials, (v) PA$8$ $-$ $8$ proton affinities, (vi) BH$6$ $-$ $6$ barrier heights, (vii) HB$6$ $-$ $6$ 
hydrogen bonding, and (viii) DI$6$ $-$ $6$ dipole bonding. Whereas, our solid-state test set contains equilibrium lattice 
constants (LC$29$), bulk moduli (BM$29$), and cohesive energies (COH$29$) of $29$ solids which includes: simple metals (Li, Na, K, Ca, 
Sr, Ba, Al), transition metals (V, Ni, Cu, Rh, Pd, Ag, Pt), ionic solids (LiF, LiCl, NaF, NaCl, MgO), insulators (BN, BP,
AlN), and semiconductors (C, Si, Ge, SiC, GaN, GaP, GaAs). All the bulk calculations are performed with the $16\times 16\times 
16$ $\Gamma$-centred $\bf{k}$ points with energy cutoff $700$ eV for the smooth convergence w.r.t. the plane-wave and energy 
convergence. All solid-state calculations are done considering non-magnetic phases and ambient-condition crystal structures
except Ni. For which the magnetic calculations are taken into account. The anti-symmetric box size of $18\times 19\times 20$ 
\AA$^3$ is considered for the atomic calculations. 

Besides all these fundamental tests, we also calculate the jellium surface XC energies and surface energies of real metallic 
systems to incorporate the functional performances. Regarding the surface energy calculations, it is done with $16 \times 16
\times 1$ $\Gamma-$ centered $\bf{k}$ points with energy convergence criterion $1.0\times 10^{-6}$ eV and energy cutoff $700$ eV. 
The dipole corrections are also taken into consideration in this calculation along with  $>$ $20$ \AA~vacuum to avoid the 
interaction between the periodic surfaces. The functionals considered for comparison with the revTM are LDA, Perdew-Burke-Ernzerhof 
(PBE)~\cite{PhysRevLett.77.3865}, PBE reparametrized for solids (PBEsol)~\cite{PhysRevLett.100.136406}, TPSS~\cite{PhysRevLett.91.146401},
revTPSS~\cite{PhysRevLett.103.026403}, Strongly Constrained and Appropriately Normed  (SCAN)~\cite{PhysRevLett.115.036402}, and TM
~\cite{PhysRevLett.117.073001}. The accuracy in the performance of all the functionals are assessed by calculating mean error (ME), 
mean absolute error (MAE), and mean absolute relative percentage error (MARPE).

\section{Results}

\subsection{Thermochemical accuracy}
Let's start with the thermochemical accuracy of each functional. The results of all tests are summarized in Table~\ref{thermo}, 
where we report the ME and MAE of the individual test set for each functionals. Inspection of Table~\ref{thermo} shows that 
the revised version of TM i.e., revTM performs better than TM for most of the thermochemical test sets. In particular, for AE$6$,
EA$13$, PA$8$, HB$6$ and DI$6$ test cases, the revTM shows improvement. However, in the atomization energies test sets AE$17$ and IP$13$, 
the performance of TM and revTM qualitatively same. The ME of IP$13$ is predicted to be much better in revTM than other functionals 
considered here. In the case of BH$6$, the TM shows better performance than revTM. It is noteworthy to mention that accurate 
prediction of barrier heights are quite difficult for semilocal functionals because of the many-electron self-interaction 
problems in the transition state of those molecules. The range-separated hybrid functionals perform better in such cases
because of the proper incorporation of the non-locality information. We also observe that for hydrogen (HB$6$) and dipole (DI$6$)
bond dissociation energies, the revTM improves the performance of TM. As those are noncovalent interactions and improvement of 
revTM over TM clearly indicates the change in correlation is important. It is also very interesting observation that, even 
though the revTPSS correlation does not improve dramatically over the TPSS correlation as far as the thermochemical accuracy 
is concerned but the change in correlation in the revTM suits perfectly for most of the thermochemical test sets. 

Now, concerning the performance of TM based functionals with other meta-GGAs and GGAs, the SCAN performs very well for the 
AE$6$ atomization energies. However, we observe that the MAE of revTM for the EA$13$ matches closely with SCAN for which
the PBE performs better than others. For IP$13$, TPSS, revTPSS, TM and revTM performs similarly. In this case, the SCAN 
functional gives MAE of $4.47$ kcal/mol. In particular, the SCAN functional is quite accurate in case of AE$6$, PA$8$, and
BH$6$ test sets. From the improvement of revTM over TM functional, it is quite clear that the change in correlation results 
to balance description of different thermochemical test sets and thus the change in correlation part in relevant. However, 
other modifications based on the $\beta$ parameter are also proposed in the TPSSloc functionals and the resultant functional 
based on the TPSSloc~\cite{PhysRevB.86.035130} is the BLOC~\cite{doi:10.1021/ct400148r} functional which also performs quite 
accurately over an wide range of molecular properties. 

\begingroup
\squeezetable
\begin{table*}
\caption{\label{tab_latt}Equilibrium lattice constants (in \AA) of different solids are obtained using several functionals
considered here. The experimental values (except AlN) are collected from references~\cite{PhysRevB.79.085104}, where the
correction due to ZPAE is taken into account. The ZPAE corrected AlN reference value is taken from ref.~\cite{PhysRevB.81.115126}. 
The best values are in bold and the most deviating values are underlined.}
{\scriptsize
\begin{ruledtabular}
\begin{tabular}{cccccccccccccccccccccccccccccccccc}
\multicolumn{1}{l}{Solids} &
\multicolumn{1}{c}{LDA} &
\multicolumn{1}{c}{PBE} &
\multicolumn{1}{c}{PBEsol} &
\multicolumn{1}{c}{TPSS} &
\multicolumn{1}{c}{revTPSS} &
\multicolumn{1}{c}{SCAN} &
\multicolumn{1}{c}{TM} &
\multicolumn{1}{c}{revTM} &
\multicolumn{1}{c}{Expt.-ZPAE}\\
\hline

\multicolumn{10}{c}{Simple metals}\\[0.2cm]
Li	&	\underline{3.366}	&	3.438	&	3.442	&	3.455	&   \textbf{3.449}	&	3.467	&	3.401	&	3.425	&	3.451	\\
Na	&	\underline{4.048}	&	4.188	&	4.167	&	4.222	&	\textbf{4.206}	&	4.180	&	4.109	&	4.149	&	4.209	\\
K	&   \underline{5.010}	&	5.247	&	\textbf{5.189}	&	5.357	&	5.325	&	5.264	&	5.140	&	5.183	&	5.212	\\
Ca	&	\underline{5.306}	&	5.502	&	5.427	&	\textbf{5.522}	&	5.515	&	5.500	&	5.453	&	5.478	&	5.556	\\
Sr	&	\underline{5.783}	&	5.996	&	5.904	&	\textbf{6.033}	&	5.996	&	6.060	&	5.955	&	5.920	&	6.040	\\
Ba	&	\underline{4.766}	&	5.020	&	4.892	&	\textbf{4.995}	&	4.972	&	5.029	&	4.971	&	4.977	&	5.002	\\
Al	&	3.978	&	4.034	&	\textbf{4.010}	&	4.008	&	4.003	&	4.001	&	\underline{3.977}	&	3.999	&	4.019	\\[0.1cm] 
	&		&		&		&		&		&		&		&		&		\\
ME	&	\underline{-0.176}	&	-0.009	&	-0.065	&	0.015	&	-0.003	&	\textbf{0.002}	&	-0.069	&	-0.051	&		\\
MAE	&	\underline{0.176}	&	\textbf{0.029}	&	0.065	&	0.032	&	0.036	&	0.031	&	0.069	&	0.051	&		\\
MARPE	&	\underline{3.522}	&	\textbf{0.569}	&	1.242	&	0.621	&	0.680	&	0.640	&	1.447	&	1.018	&		\\[0.2cm]

\multicolumn{10}{c}{Transition metals}\\[0.2cm]
V	&	\underline{2.911}	&	\textbf{2.978}	&	2.944	&	2.957	&	2.951	&	2.953	&	2.960	&	2.954	&	3.024	\\
Ni	&	\underline{3.419}	&	\textbf{3.516}	&	3.459	&	3.470	&	3.454	&	3.465	&	3.449	&	3.443	&	3.508	\\
Cu	&	\underline{3.519}	&	3.634	&	3.565	&	\textbf{3.569}	&	3.540	&	3.556	&	3.530	&	3.534	&	3.596	\\
Rh	&	\underline{3.749}	&	3.822	&	3.772	&	3.802	&	3.779	&	3.782	&	\textbf{3.786}	&	3.785	&	3.793	\\
Pd	&	3.837	&	\underline{3.941}	&	3.870	&	3.905	&	\textbf{3.878}	&	3.897	&	3.887	&	3.886	&	3.876	\\
Ag	&	3.994	&	\underline{4.142}	&	4.045	&	4.085	&	4.053	&	4.076	&	\textbf{4.061}	&	\textbf{4.061}	&	4.062	\\
Pt	&	3.893	&	\underline{3.966}	&	3.914	&	3.944	&	\textbf{3.913}	&	3.897	&	3.916	&	3.917	&	3.913	\\[0.1cm] 
	&		&		&		&		&		&		&		&		&		\\
ME	&	\underline{-0.064}	&	0.032	&	-0.029	&	\textbf{-0.006}	&	-0.029	&	-0.021	&	-0.026	&	-0.027	&		\\
MAE	&	\underline{0.064}	&	0.046	&	\textbf{0.029}	&	0.032	&	0.030	&	0.031	&	0.030	&	0.031	&		\\
MARPE	&	\underline{1.824}	&	1.224	&	\textbf{0.865}	&	0.913	&	0.879	&	0.896	&	0.886	&	0.927	&		\\[0.2cm]

\multicolumn{10}{c}{Ionic solids}\\[0.2cm]
LiF	&	3.939	&	\underline{4.055}	&	4.002	&	4.014	&	4.000	&	3.972	&	\textbf{3.967}	&	3.997	&	3.960	\\
LiCl	&	\underline{4.975}	&	5.145	&	\textbf{5.072}	&	5.118	&	5.107	&	5.096	&	5.045	&	5.089	&	5.072	\\
NaF	    &	\underline{4.435}	&	4.618	&	4.546	&	4.595	&	\textbf{4.562}	&	4.476	&	4.491	&	4.534	&	4.576	\\
NaCl	&	\underline{5.430}	&	5.649	&	5.552	&	5.646	&	5.616	&	5.525	&	5.500	&	\textbf{5.558}	&	5.565	\\
MgO	&	4.146	&	\underline{4.246}	&	4.209	&	4.228	&	4.226	&	\textbf{4.185}	&	4.205	&	4.219	&	4.186	\\[0.1cm] 
	&		&		&		&		&		&		&		&		&		\\
ME	&	\underline{-0.087}	&	0.071	&	\textbf{0.004}	&	0.048	&	0.030	&	-0.021	&	-0.030	&	0.008	&		\\
MAE	&	\underline{0.087}	&	0.071	&	\textbf{0.022}	&	0.048	&	0.036	&	0.035	&	0.041	&	0.027	&		\\
MARPE	&	\underline{1.781}	&	1.540	&	\textbf{0.500}	&	1.029	&	0.776	&	0.741	&	0.838	&	0.620	&		\\[0.2cm]

\multicolumn{10}{c}{Insulators}\\[0.2cm]

BN	&	\textbf{3.584}	&	\underline{3.625}	&	3.608	&	3.625	&	3.619	&	3.605	&	3.608	&	3.611	&	3.585	\\
BP	&	\underline{4.490}	&	4.547	&	4.522	&	4.545	&	4.531	&	4.521	&	4.511	&	\textbf{4.519}	&	4.520	\\
AlN	&	4.346	&	\underline{4.401}	&	4.377	&	4.386	&	4.382	&	4.361	&	\textbf{4.367}	&	4.378	&	4.368	\\[0.1cm] 
	&		&		&		&		&		&		&		&		&		\\
ME	&	-0.018	&	\underline{0.033}	&	0.011	&	0.028	&	0.020	&	0.005	&	\textbf{0.004}	&	0.012	&		\\
MAE	&	0.018	&	\underline{0.033}	&	0.011	&	0.028	&	0.020	&	\textbf{0.009}	&	0.011	&	0.012	&		\\
MARPE	&	0.398	&	\underline{0.823}	&	0.297	&	0.694	&	0.504	&	\textbf{0.247}	&	0.288	&	0.325	&		\\[0.2cm]

\multicolumn{10}{c}{Semiconductors}\\[0.2cm]
C	&	\textbf{3.536}	&	\underline{3.574}	&	3.557	&	3.572	&	3.563	&	3.556	&	3.555	&	3.556	&	3.544	\\
Si	&	5.402	&	\underline{5.466}	&	5.434	&	5.449	&	5.436	&	5.426	&	\textbf{5.411}	&	5.448	&	5.415	\\
Ge	&	\textbf{5.646}	&	\underline{5.782}	&	5.702	&	5.752	&	5.709	&	5.685	&	5.671	&	5.689	&	5.639	\\
SiC	&	4.332	&	\underline{4.379}	&	4.359	&	4.365	&	4.357	&	4.352	&	\textbf{4.344}	&	4.351	&	4.340	\\
GaN	&	4.503	&	\underline{4.588}	&	4.547	&	4.581	&	4.570	&	\textbf{4.526}	&	4.550	&	4.559	&	4.520	\\
GaP	&	\textbf{5.425}	&	\underline{5.533}	&	5.474	&	5.523	&	5.500	&	5.456	&	5.464	&	5.480	&	5.435	\\
GaAs	&	\textbf{5.626}	&	\underline{5.762}	&	5.683	&	5.736	&	5.699	&	5.665	&	5.662	&	5.680	&	5.637	\\[0.1cm] 
	&		&		&		&		&		&		&		&		&		\\
ME	&	\textbf{-0.009}	&	\underline{0.079}	&	0.032	&	0.064	&	0.043	&	0.019	&	0.018	&	0.033	&		\\
MAE	&	\textbf{0.011}	&	\underline{0.079}	&	0.032	&	0.064	&	0.043	&	0.019	&	0.019	&	0.033	&		\\
MARPE	&	\textbf{0.219}	&	\underline{1.535}	&	0.629	&	1.246	&	0.851	&	0.379	&	0.384	&	0.649	&		\\[0.2cm]

\hline
TME	&	\underline{-0.089}	&	0.040	&	-0.013	&	0.029	&	0.010	&	\textbf{-0.003}	&	-0.023	&	-0.008	\\
TMAE	&	\underline{0.089}	&	0.053	&	0.036	&	0.042	&	0.034	&	\textbf{0.027}	&	0.037	&	0.034	\\
TMARPE	&	\underline{1.952}	&	1.154	&	0.778	&	0.920	&	0.768	&	\textbf{0.615}	&	0.830	&	0.767	\\

	\end{tabular}
\end{ruledtabular}
}
\end{table*}
\endgroup

\begin{figure*}
\begin{center}
\includegraphics[width=6.0in,height=2.5in,angle=0.0]{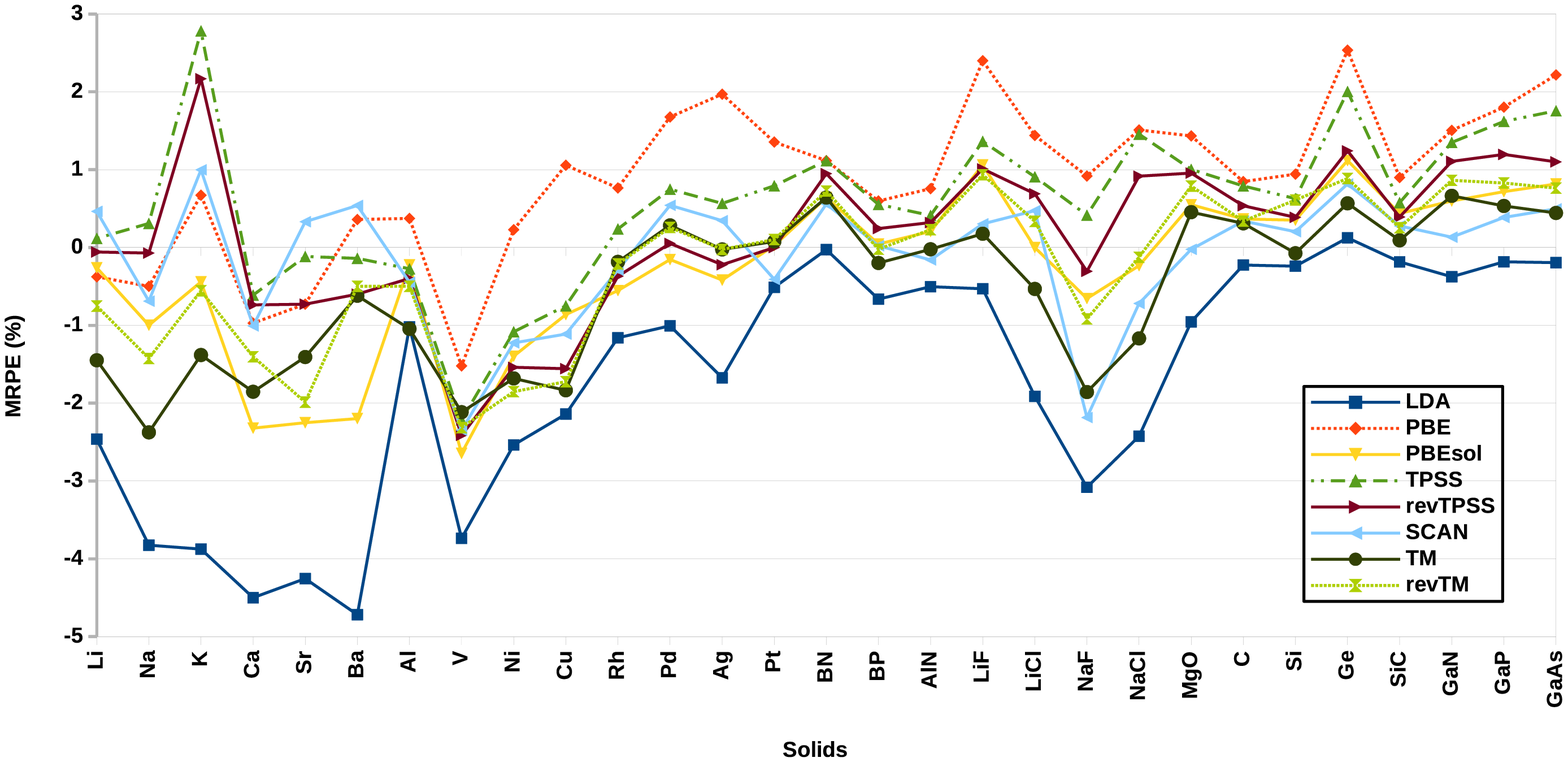} 
\end{center}
\caption{Shown is the MARE of the lattice constants obtained from different functionals presented in Table~\ref{tab_latt}}
\label{figaax}
\end{figure*}

\begingroup
\squeezetable
\begin{table*}
\caption{\label{tab_bm} Bulk moduli (in GPa) of different solids are obtained using concerned functionals. The experimental 
values are taken from ref.~\cite{PhysRevB.84.035117,doi:10.1063/1.5047863,doi:10.1063/1.4948636,PhysRevB.81.115126}. The best 
values are in bold and the most deviating values are underlined.}
{\scriptsize
\begin{ruledtabular}
\begin{tabular}{cccccccccccccccccccccccccccccccccc}
\multicolumn{1}{l}{Solids} &
\multicolumn{1}{c}{LDA} &
\multicolumn{1}{c}{PBE} &
\multicolumn{1}{c}{PBEsol} &
\multicolumn{1}{c}{TPSS} &
\multicolumn{1}{c}{revTPSS} &
\multicolumn{1}{c}{SCAN} &
\multicolumn{1}{c}{TM} &
\multicolumn{1}{c}{revTM} &
\multicolumn{1}{c}{Expt.}\\
\hline

\multicolumn{10}{c}{Simple metals}\\[0.2cm]
Li	&	\underline{15.3}	&	14.1	&	\textbf{13.8}	&	13.6	&	13.6	&	13.3	&	14.7	&	14.1	&	13.9	\\
Na	&	\underline{9.3}	&	8.0	&	\textbf{7.9}	&	7.4	&	\textbf{7.5}	&	\textbf{7.9}	&	8.9	&	8.6	&	7.7	\\
K	&	\underline{4.3}	&	3.5	&	3.5	&	3.3	&	3.3	&	3.2	&	3.8	&	\textbf{3.7}	&	3.7	\\
Ca	&	17.9	&	16.8	&	17.2	&	\underline{16.8}	&	17.1	&	17.6	&	\textbf{18.5}	&	17.9	&	18.7	\\
Sr	&	\underline{14.2}	&	11.5	&	12.3	&	11.4	&	11.7	&	11.1	&	\textbf{12.6}	&	\textbf{12.4}	&	12.5	\\
Ba	&	10.6	&	8.7	&	\textbf{9.3}	&	8.4	&	8.7	&	\underline{8.1}	&	9.2	&	\textbf{9.3}	&	9.4	\\
Al	&	80.6	&	74.3	&	78.9	&	\textbf{82.2}	&	82.9	&	79.8	&	\underline{89.8}	&	84.1	&	82.0	\\[0.1cm] 
	&		&		&		&		&		&		&		&		&		\\
ME	&	0.6	&	\underline{-1.6}	&	-0.7	&	-0.7	&	-0.4	&	-1.0	&	1.4	&	\textbf{0.3}	&		\\
MAE	&	1.2	&	\underline{1.7}	&	0.8	&	0.7	&	0.7	&	1.0	&	1.5	&	\textbf{0.6}	&		\\
MARPE	&	\underline{11.3}	&	6.5	&	3.3	&	6.7	&	5.6	&	7.6	&	5.3	&	\textbf{3.1}	&		\\[0.2cm]

\multicolumn{10}{c}{Transition metals}\\[0.2cm]
V	&	\underline{206.2}	&	187.8	&	204.0	&	201.8	&	205.6	&	203.8	&	\textbf{181.8}	&	193.6	&	158.9	\\
Ni	&	\underline{250.3}	&	\textbf{208.6}	&	231.9	&	228.7	&	244.7	&	241.5	&	238.2	&	243.5	&	185.0	\\
Cu	&	\underline{182.9}	&	137.1	&	163.3	&	156.5	&	170.5	&	\textbf{152.4}	&	164.2	&	164.6	&	145.0	\\
Rh	&	\underline{311.4}	&	254.8	&	295.8	&	\textbf{276.7}	&	292.4	&	290.3	&	284.0	&	281.3	&	272.1	\\
Pd	&	222.6	&	\underline{165.3}	&	201.8	&	187.9	&	201.1	&	193.5	&	\textbf{195.3}	&	194.9	&	198.1	\\
Ag	&	\underline{135.8}	&	86.1	&	\textbf{112.3}	&	102.3	&	113.0	&	105.4	&	113.0	&	116.1	&	110.8	\\
Pt	&	302.0	&	245.0	&	285.8	&	264.6	&	\textbf{284.1}	&	\underline{244.1}	&	280.4	&	279.7	&	284.2	\\[0.1cm] 
	&		&		&		&		&		&		&		&		&		\\
ME	&	\underline{36.7}	&	-9.9	&	20.1	&	\textbf{9.2}	&	22.5	&	11.0	&	14.7	&	17.1	&		\\
MAE	&	\underline{36.7}	&	24.9	&	20.1	&	20.1	&	22.5	&	25.3	&	\textbf{16.6}	&	19.3	&		\\
MARPE	&	\underline{21.0}	&	13.6	&	11.3	&	11.4	&	12.9	&	13.1	&	\textbf{9.4}	&	11.2	&		\\[0.2cm]

\multicolumn{10}{c}{Ionic solids}\\[0.2cm]
LiF	&	\underline{88.3}	&	68.2	&	73.4	&	67.4	&	70.0	&	81.3	&	80.4	&	\textbf{74.8}	&	76.3	\\
LiCl	&	41.2	&	\underline{31.7}	&	34.8	&	32.2	&	33.1	&	35.6	&	\textbf{37.2}	&	34.5	&	38.7	\\
NaF	&	\underline{66.1}	&	47.9	&	\textbf{52.5}	&	46.3	&	49.1	&	62.7	&	60.1	&	54.3	&	53.1	\\
NaCl	&	\underline{33.1}	&	24.5	&	26.8	&	23.6	&	24.9	&	30.1	&	30.9	&	\textbf{27.6}	&	27.6	\\
MgO	&	179.5	&	\underline{151.8}	&	160.8	&	157.0	&	157.6	&	\textbf{173.6}	&	165.3	&	159.8	&	169.8	\\[0.1cm] 
	&		&		&		&		&		&		&		&		&		\\
ME	&	\underline{8.5}	&	-8.3	&	-3.4	&	-7.8	&	-6.2	&	3.6	&	\textbf{1.7}	&	-2.9	&		\\
MAE	&	\underline{8.5}	&	8.3	&	3.4	&	7.8	&	6.2	&	4.8	&	4.1	&	\textbf{3.4}	&		\\
MARPE	&	\underline{14.5}	&	12.1	&	4.6	&	12.7	&	9.4	&	8.8	&	7.4	&	\textbf{4.2}	&		\\[0.2cm]

\multicolumn{10}{c}{Insulators}\\[0.2cm]
BN	&	\textbf{398.2}	&	\underline{368.4}	&	382.1	&	370.7	&	373.8	&	393.7	&	384.2	&	380.5	&	410.2	\\
BP	&	173.2	&	159.7	&	166.9	&	\underline{159.2}	&	162.2	&	171.3	&	170.2	&	\textbf{168.2}	&	168.0	\\
AlN	&	208.5	&	\underline{191.1}	&	199.1	&	198.3	&	199.0	&	210.2	&	205.1	&	\textbf{200.1}	&	202.0	\\[0.1cm] 
	&		&		&		&		&		&		&		&		&		\\
ME	&	\textbf{-0.1}	&	\underline{-20.3}	&	-10.7	&	-17.3	&	-15.1	&	-1.7	&	-6.9	&	-10.5	&		\\
MAE	&	\textbf{7.9}	&	\underline{20.3}	&	10.7	&	17.3	&	15.1	&	9.3	&	10.4	&	10.6	&		\\
MARPE	&	3.1	&	\underline{6.8}	&	3.0	&	5.6	&	4.6	&	3.3	&	3.1	&	\textbf{2.8}	&		\\[0.2cm]

\multicolumn{10}{c}{Semiconductors}\\[0.2cm]
C	&	459.5	&	425.2	&	442.5	&	\underline{421.3}	&	430.5	&	\textbf{450.3}	&	444.8	&	441.6	&	454.7	\\
Si	&	94.5	&	\underline{86.7}	&	91.6	&	90.0	&	91.8	&	97.2	&	\textbf{97.3}	&	94.0	&	100.8	\\
Ge	&	71.3	&	\underline{58.3}	&	66.4	&	60.8	&	65.1	&	71.1	&	\textbf{72.1}	&	69.4	&	77.3	\\
SiC	&	\textbf{224.9}	&	\underline{208.1}	&	217.0	&	213.6	&	217.6	&	222.5	&	224.1	&	220.3	&	229.1	\\
GaN	&	\textbf{209.4}	&	\underline{178.5}	&	194.5	&	184.2	&	187.4	&	209.0	&	197.0	&	192.2	&	210.0	\\
GaP	&	90.1	&	\underline{77.0}	&	84.7	&	78.7	&	81.6	&	89.8	&	\textbf{88.8}	&	85.7	&	89.0	\\
GaAs	&	\textbf{75.1}	&	\underline{61.3}	&	69.6	&	63.9	&	68.0	&	74.2	&	74.5	&	71.6	&	76.7	\\[0.1cm] 
	&		&		&		&		&		&		&		&		&		\\
ME	&	\textbf{-1.8}	&	\underline{-20.4}	&	-10.2	&	-17.9	&	-13.7	&	-3.4	&	-5.6	&	-9.0	&		\\
MAE	&	\textbf{3.5}	&	\underline{20.4}	&	10.2	&	17.9	&	13.7	&	3.6	&	5.6	&	9.0	&		\\
MARPE	&	\textbf{2.9}	&	\underline{14.7}	&	7.5	&	12.4	&	9.4	&	2.9	&	3.4	&	6.1	&		\\[0.2cm]

\hline
TME	&	10.0	&	\underline{-11.2}	&	\textbf{0.5}	&	-5.4	&	-0.6	&	2.0	&	2.1	&	0.5	&		\\
TMAE	&	12.3	&	\underline{14.9}	&	9.2	&	12.5	&	11.5	&	9.0	&	\textbf{7.5}	&	8.6	&		\\
TMARPE	&	\underline{11.3}	&	11.2	&	6.4	&	10.1	&	8.8	&	7.6	&	6.0	&	\textbf{5.9}	&		\\

	\end{tabular}
\end{ruledtabular}
}
\end{table*}
\endgroup

\begin{figure*}
\begin{center}
\includegraphics[width=6.0in,height=2.5in,angle=0.0]{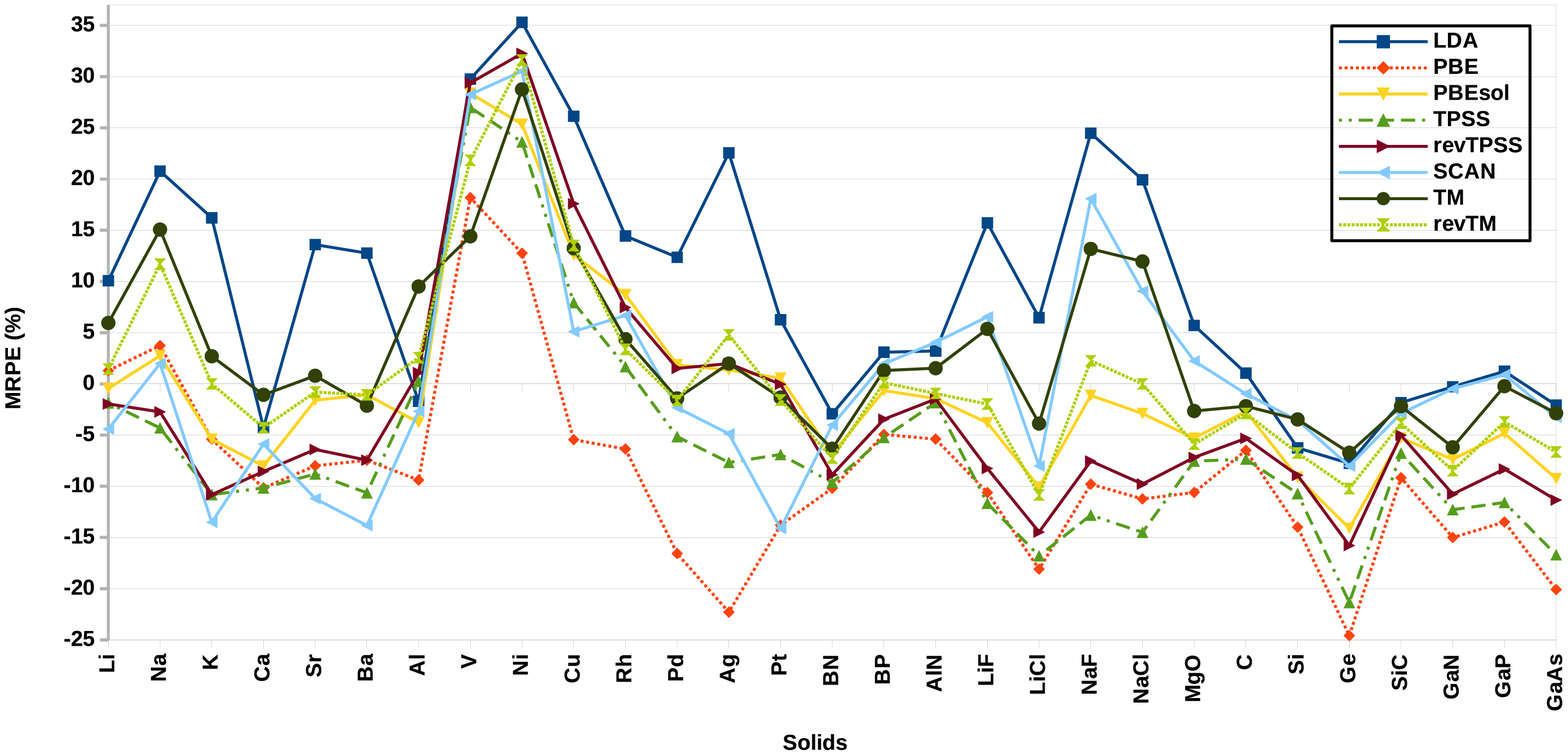} 
\end{center}
\caption{Shown is the MARE of the bulk moduli of different functionals presented in Table~\ref{tab_bm}}
\label{fig5}
\end{figure*}

\begingroup
\squeezetable
\begin{table*}
\caption{\label{tab_coh}Cohesive energies (in eV/atom) of different solids are obtained using different functionals. The 
experimental values are taken from ref.~\cite{PhysRevB.84.035117,doi:10.1063/1.5047863,doi:10.1063/1.4948636,
PhysRevB.81.115126}. The best values are in bold and the most deviating values are underlined.}
{\scriptsize
\begin{ruledtabular}
\begin{tabular}{cccccccccccccccccccccccccccccccccc}
\multicolumn{1}{l}{Solids} &
\multicolumn{1}{c}{LDA} &
\multicolumn{1}{c}{PBE} &
\multicolumn{1}{c}{PBEsol} &
\multicolumn{1}{c}{TPSS} &
\multicolumn{1}{c}{revTPSS} &
\multicolumn{1}{c}{SCAN} &
\multicolumn{1}{c}{TM} &
\multicolumn{1}{c}{revTM} &
\multicolumn{1}{c}{Expt.}\\
\hline

\multicolumn{10}{c}{Simple metals}\\[0.2cm]
Li	&	\underline{1.807}	&	1.599	&	\textbf{1.669}	&	1.631	&	1.638	&	1.562	&	1.676	&	1.630	&	1.658	\\
Na	&	\underline{1.254}	&	1.083	&	1.154	&	\textbf{1.140}	&	1.159	&	1.097	&	1.215	&	1.144	&	1.119	\\
K	&	1.028	&	0.865	&	0.924	&	0.918	&	\textbf{0.941}	&	\underline{0.835}	&	0.987	&	0.931	&	0.940	\\
Ca	&	2.222	&	\textbf{1.913}	&	2.112	&	2.025	&	2.075	&	2.081	&	\underline{2.287}	&	2.062	&	1.860	\\
Sr	&	1.896	&	1.610	&	1.809	&	\textbf{1.756}	&	1.827	&	1.821	&	\underline{2.062}	&	1.935	&	1.730	\\
Ba	&	2.249	&	\textbf{1.882}	&	2.119	&	2.027	&	2.106	&	2.037	&	\underline{2.336}	&	2.197	&	1.910	\\
Al	&	3.899	&	\textbf{3.389}	&	3.742	&	3.513	&	3.635	&	3.607	&	\underline{3.917}	&	3.771	&	3.431	\\[0.1cm] 
	&		&		&		&		&		&		&		&		&		\\
ME	&	0.244	&	\textbf{-0.044}	&	0.126	&	0.052	&	0.105	&	0.056	&	\underline{0.262}	&	0.146	&		\\
MAE	&	0.244	&	\textbf{0.059}	&	0.130	&	0.066	&	0.110	&	0.120	&	\underline{0.262}	&	0.157	&		\\
MARPE	&	12.980	&	3.890	&	6.231	&	\textbf{3.534}	&	5.466	&	6.835	&	\underline{13.326}	&	7.504	&		\\[0.2cm]

\multicolumn{10}{c}{Transition metals}\\[0.2cm]
V	&	\underline{6.633}	&	\textbf{5.251}	&	5.831	&	5.509	&	5.760	&	4.965	&	5.933	&	5.811	&	5.340	\\
Ni	&	\underline{6.216}	&	\textbf{4.677}	&	5.339	&	5.067	&	5.441	&	5.251	&	5.692	&	5.535	&	4.480	\\
Cu	&	\underline{4.528}	&	\textbf{3.484}	&	4.029	&	3.749	&	4.092	&	3.873	&	4.384	&	4.286	&	3.524	\\
Rh	&	\underline{7.417}	&	\textbf{5.855}	&	6.731	&	6.001	&	6.327	&	5.582	&	6.549	&	6.691	&	5.783	\\
Pd	&	\underline{5.065}	&	3.741	&	4.470	&	\textbf{4.002}	&	4.395	&	4.383	&	4.710	&	4.632	&	3.938	\\
Ag	&	\underline{3.637}	&	2.520	&	3.075	&	2.728	&	\textbf{3.041}	&	2.883	&	3.351	&	3.248	&	2.985	\\
Pt	&	\underline{6.943}	&	5.416	&	6.267	&	\textbf{5.741}	&	6.189	&	6.172	&	6.459	&	6.350	&	5.870	\\[0.1cm] 
	&		&		&		&		&		&		&		&		&		\\
ME	&	\underline{1.217}	&	-0.139	&	0.546	&	\textbf{0.125}	&	0.475	&	0.170	&	0.737	&	0.662	&		\\
MAE	&	\underline{1.217}	&	\textbf{0.216}	&	0.546	&	0.236	&	0.475	&	0.364	&	0.737	&	0.662	&		\\
MARPE	&	\underline{26.921}	&	\textbf{5.251}	&	11.769	&	5.551	&	10.537	&	8.211	&	16.815	&	14.901	&		\\[0.2cm]

\multicolumn{10}{c}{Ionic solids}\\[0.2cm]
LiF	&	\underline{4.847}	&	4.364	&	4.497	&	4.341	&	4.284	&	4.708	&	4.520	&	\textbf{4.428}	&	4.457	\\
LiCl	&	3.752	&	\underline{3.345}	&	3.480	&	3.401	&	3.442	&	3.439	&	\textbf{3.546}	&	3.451	&	3.586	\\
NaF	&	\underline{4.381}	&	3.925	&	4.049	&	3.910	&	3.859	&	4.324	&	4.116	&	\textbf{4.009}	&	3.970	\\
NaCl	&	3.454	&	\underline{3.102}	&	3.214	&	3.169	&	3.215	&	3.249	&	\textbf{3.340}	&	3.225	&	3.337	\\
MgO	&	\underline{5.947}	&	5.010	&	5.351	&	4.945	&	4.877	&	5.413	&	\textbf{5.214}	&	5.079	&	5.203	\\[0.1cm] 
	&		&		&		&		&		&		&		&		&		\\
ME	&	\underline{0.366}	&	-0.161	&	\textbf{0.008}	&	-0.157	&	-0.175	&	0.116	&	0.037	&	-0.072	&		\\
MAE	&	\underline{0.366}	&	0.161	&	0.099	&	0.157	&	0.175	&	0.210	&	\textbf{0.053}	&	0.088	&		\\
MARPE	&	\underline{8.308}	&	4.138	&	2.475	&	3.853	&	4.123	&	5.064	&	\textbf{1.302}	&	2.227	&		\\[0.2cm]

\multicolumn{10}{c}{Insulators}\\[0.2cm]
BN	&	\underline{8.073}	&	6.934	&	7.394	&	6.635	&	6.639	&	\textbf{6.839}	&	6.970	&	6.903	&	6.760	\\
BP	&	\underline{6.249}	&	5.286	&	5.717	&	5.027	&	\textbf{5.070}	&	5.309	&	5.405	&	5.325	&	5.140	\\
AlN	&	\underline{6.603}	&	5.691	&	6.045	&	5.685	&	5.721	&	\textbf{5.844}	&	5.963	&	5.840	&	5.850	\\[0.1cm] 
	&		&		&		&		&		&		&		&		&		\\
ME	&	\underline{1.058}	&	\textbf{0.054}	&	0.469	&	-0.134	&	-0.107	&	0.081	&	0.196	&	0.106	&		\\
MAE	&	\underline{1.058}	&	0.160	&	0.469	&	0.134	&	0.107	&	\textbf{0.085}	&	0.196	&	0.113	&		\\
MARPE	&	\underline{17.957}	&	2.711	&	7.979	&	2.289	&	1.786	&	\textbf{1.520}	&	3.398	&	1.962	&		\\[0.2cm]

\multicolumn{10}{c}{Semiconductors}\\[0.2cm]
C	&	\underline{8.842}	&	7.697	&	8.198	&	7.392	&	7.312	&	\textbf{7.513}	&	7.632	&	7.586	&	7.545	\\
Si	&	\underline{5.191}	&	4.461	&	4.806	&	4.440	&	4.525	&	4.808	&	4.783	&	\textbf{4.677}	&	4.685	\\
Ge	&	\underline{4.460}	&	3.590	&	3.980	&	3.593	&	3.755	&	4.024	&	4.003	&	\textbf{3.924}	&	3.918	\\
SiC	&	\underline{7.298}	&	6.352	&	6.775	&	6.287	&	6.288	&	6.501	&	6.550	&	\textbf{6.461}	&	6.478	\\
GaN	&	\underline{5.382}	&	4.352	&	4.762	&	4.290	&	4.398	&	4.470	&	4.698	&	\textbf{4.590}	&	4.550	\\
GaP	&	\underline{4.308}	&	3.438	&	3.827	&	3.408	&	3.538	&	\textbf{3.640}	&	3.803	&	3.705	&	3.610	\\
GaAs	&	\underline{4.020}	&	3.122	&	3.512	&	3.122	&	3.269	&	\textbf{3.352}	&	3.512	&	3.431	&	3.337	\\[0.1cm] 
	&		&		&		&		&		&		&		&		&		\\
ME	&	\underline{0.768}	&	-0.159	&	0.248	&	-0.227	&	-0.148	&	\textbf{0.026}	&	0.123	&	0.036	&		\\
MAE	&	\underline{0.768}	&	0.202	&	0.248	&	0.227	&	0.148	&	0.058	&	0.123	&	\textbf{0.043}	&		\\
MARPE	&	\underline{16.082}	&	4.667	&	4.760	&	5.179	&	2.996	&	1.307	&	2.910	&	\textbf{1.065}	&		\\[0.2cm]

\hline
TME	&	\underline{0.711}	&	-0.105	&	0.272	&	\textbf{-0.053}	&	0.063	&	0.089	&	0.297	&	0.202	\\
TMAE	&	\underline{0.711}	&	\textbf{0.160}	&	0.289	&	0.169	&	0.218	&	0.176	&	0.300	&	0.235	\\
TMARPE	&	\underline{16.803}	&	\textbf{4.327}	&	6.746	&	4.344	&	5.481	&	4.978	&	8.554	&	6.252	\\

	\end{tabular}
\end{ruledtabular}
}
\end{table*}
\endgroup

\begin{figure*}
\begin{center}
\includegraphics[width=6.0in,height=2.5in,angle=0.0]{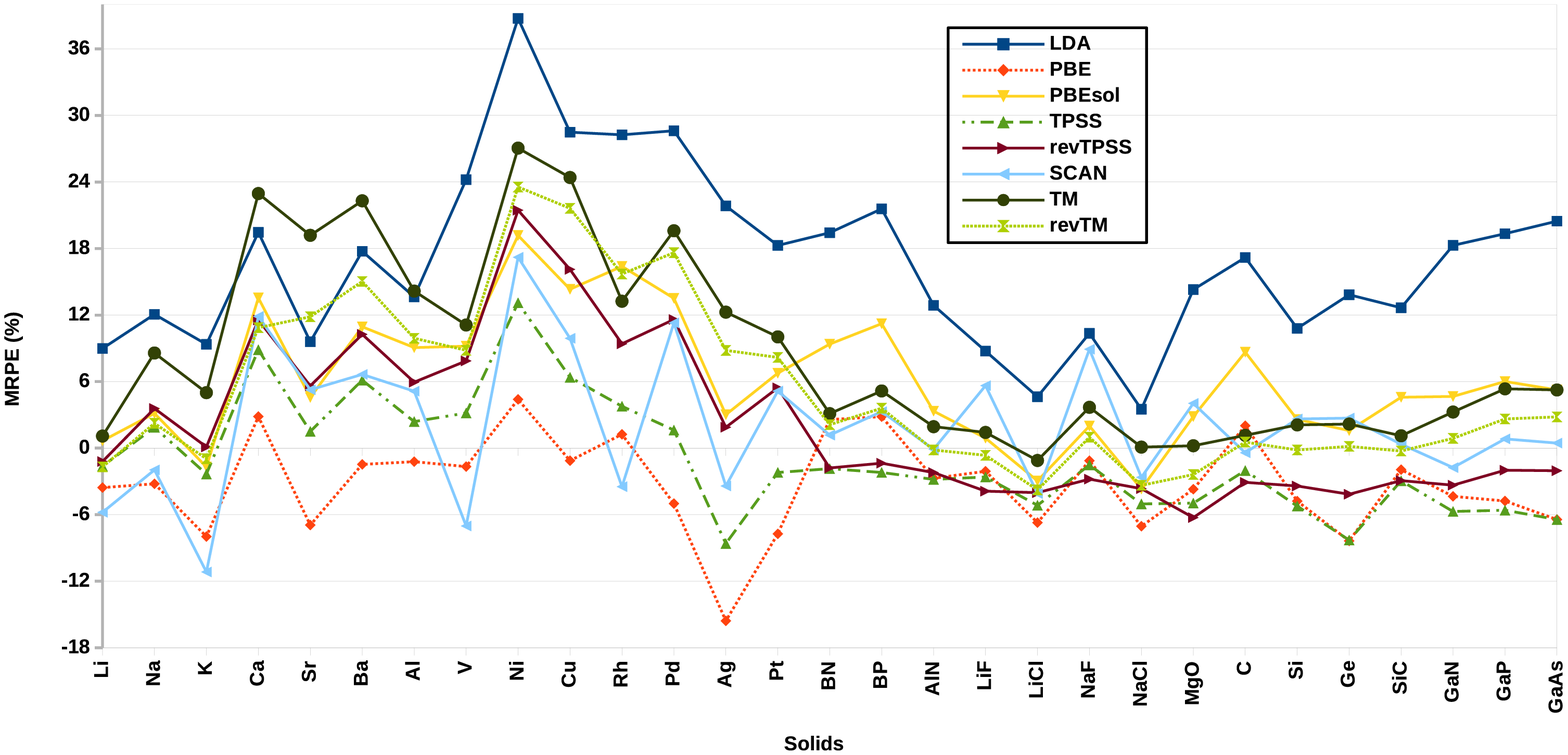} 
\end{center}
\caption{Shown is the MARE of the cohesive energies of different functionals presented in Table~\ref{tab_coh}}
\label{fig6}
\end{figure*}

\subsection{Solid-state performances}
Having established the thermochemical accuracy of the revTM functional, we now focus on the solid-state performance 
of the functional. Unlike the thermochemical test set, in the present solid-state performance, we include the PBEsol
functional which is quite accurate for the solid-state structural properties and its comparison with other functionals 
is necessary to check the robustness of the functionals performances in a more competitive manner.   

\textbf{Lattice constants :}~First, we focus on the performance in case of lattice constants. The lattice constant is 
one of the fundamental properties of solids and the accuracy of several solid-state structural properties depend on the 
accuracy of obtaining the same. In Table~\ref{tab_latt}, we summarize the performances of all the functionals. The 
zero-point anharmonic expansion (ZPAE) corrected experimental lattice constants values are taken from different 
literatures and the same are referred in the Table~\ref{tab_latt}. In Fig.~\ref{figaax}, we also plot the percentage 
deviation of the different solids given in Table~\ref{tab_latt}. 

Concerning the overall performance of the relevant functionals, as usual, the LDA functional underestimates the lattice 
constants, whereas PBE overestimates the lattice constants. The overestimation of the PBE functional is reduced by 
restoring the exact second-order gradient correction as applied in the PBEsol functional. Regarding the meta-GGA 
functionals, the TPSS functional overestimates the lattice constants which is improved by the revised version of
the TPSS i.e., revTPSS. As mentioned erlier, the revTPSS functional has been proposed by modifying the exchange 
enhancement factor such that the slowly varying density gradient approximation is satisfied for an wide range of 
the reduced density gradient ($s$). Also, modification in the correlation energy functional is incorporated in order
to take care the full linear response. Regarding the most advanced meta-GGA functionals, the SCAN improves the
meta-GGA performance and results the least MAE. Concerning the overall performances of TM and revTM functionals, the 
revTM slightly improves the performance of TM. 

Now, we focus on the performance of TM and revTM for the individual group of solid-state structures as mentioned in 
Table~\ref{tab_latt}. For the simple metals and ionic insulators, we observe a clear improvement in the lattice
constants by revTM functional. The TM functional only performs better than revTM for the semiconductors. However, for 
transition metals, and insulators both TM and revTM performs almost equivalently. The systematic improvement in the
performances of the revTM can be understood as the change in the correlation functional and the change of $\tilde{q}
\to \tilde{q}_b$ which slightly reduces the effect of exchange enhancement factor. More physically, in the intershell
region ( where  $\alpha$ may be larger even for small $s$ (as shown in Fig.~\ref{figaa})), the $F_x^{TM-sc}$ enhances 
the exchange energy density. Thus, the exchange hole becomes more centered in the intershell region than the outer
shell region which effects in the reduction in the lattice constants. On the other hand, the modification of $\tilde{q}
\to \tilde{q}_b$ in revTM functional actually reduces this effect and shows extension in the lattice constants. This
is clearly evident from Table~\ref{tab_latt}, where we observe that the lattice constants obtained using revTM functional 
for simple metals, and ionic solids extend little bit compared to its TM functional counterpart. Therefore, we can
say that this small modification in the exchange enhancement factor quite reasonably improves the lattice constants of 
those solids for which TM has the tendency to underestimates the lattice constants slightly. Concerning the correlation 
effect, the full linear response $\beta$ affects the lattice constants by cancelling the error of exchange and correlation 
in the low-density limit. However, we observe a PBEsol like tendency in the performance of revTM. Note that in the alkali 
metals and ionic solids, the long-range van der Waals interaction between the semi-core states shrink the lattice constants, 
which is correctly captured by the SCAN, TM and revTM functionals because SCAN includes the intermediate vdW interaction 
and TM based functionals include the long-range vdW interaction due to the oscillation of the exchange hole. However, 
slight elongation in the revTM is observed compared to TM because the revTM exchange enhancement factor is now de-enhanced 
than TM in the density overlap region as shown in Fig.~\ref{figaa}. Through the modifications of TM functional we observe
small but systematic improvement in the lattice constants of revTM over TM.    

\begin{table*}
\centering
\caption{The jellium surface exchange-correlation energies ($\sigma_{xc}$) (in erg/cm$^2$) of different functionals. The 
LDA, PBE, PBEsol, revTPSS values are taken from ref.~\cite{PhysRevLett.103.026403}. The SCAN values are taken from
ref.~\cite{PatraE9188}. The DMC values are taken from ref.~\cite{PatraE9188}. Rest of the functional values are calculated 
in this work.}
\begin{tabular}{lcccccccccccccccccc}
\hline
\hline
                $r_s$&$\sigma_{xc}^{LDA}$&$\sigma_{xc}^{PBE}$&$\sigma_{xc}^{PBEsol}$
                &$\sigma_{xc}^{TPSS}$&$\sigma_{xc}^{revTPSS}$&$\sigma_{xc}^{SCAN}$            &$\sigma_{xc}^{TM}$&$\sigma_{xc}^{revTM}$& DMC \\
                \hline
                $2$&3354&3265&3374&3380&3428&3422&3517&3468&3392$\pm$50\\
                $3$&764&741&774&772&783&788&824&803& 768$\pm$10 \\
                $4$&261&252&267&266&268&274&291&280&261$\pm$8\\
                $6$&53.0&52.0&56.7&55.5&55.4&58.9&64.7&60.3&53.0\\
\hline
ME&-10.5&-41.0&-0.6&-0.1&15.1 &17.2&55.7&34.3&\\
MAE&10.5&41.0&8.4&5.9&15.1 &17.2&55.7&34.3&           \\
MARE&0.4&3.1&2.6&1.9&2.5 &4.9&11.1&6.9&               \\
\hline\hline
\label{tab5}
\end{tabular}
\end{table*}

  \begin{figure*}
\begin{center}
\includegraphics[width=2.2in,height=2.2in,angle=0.0]{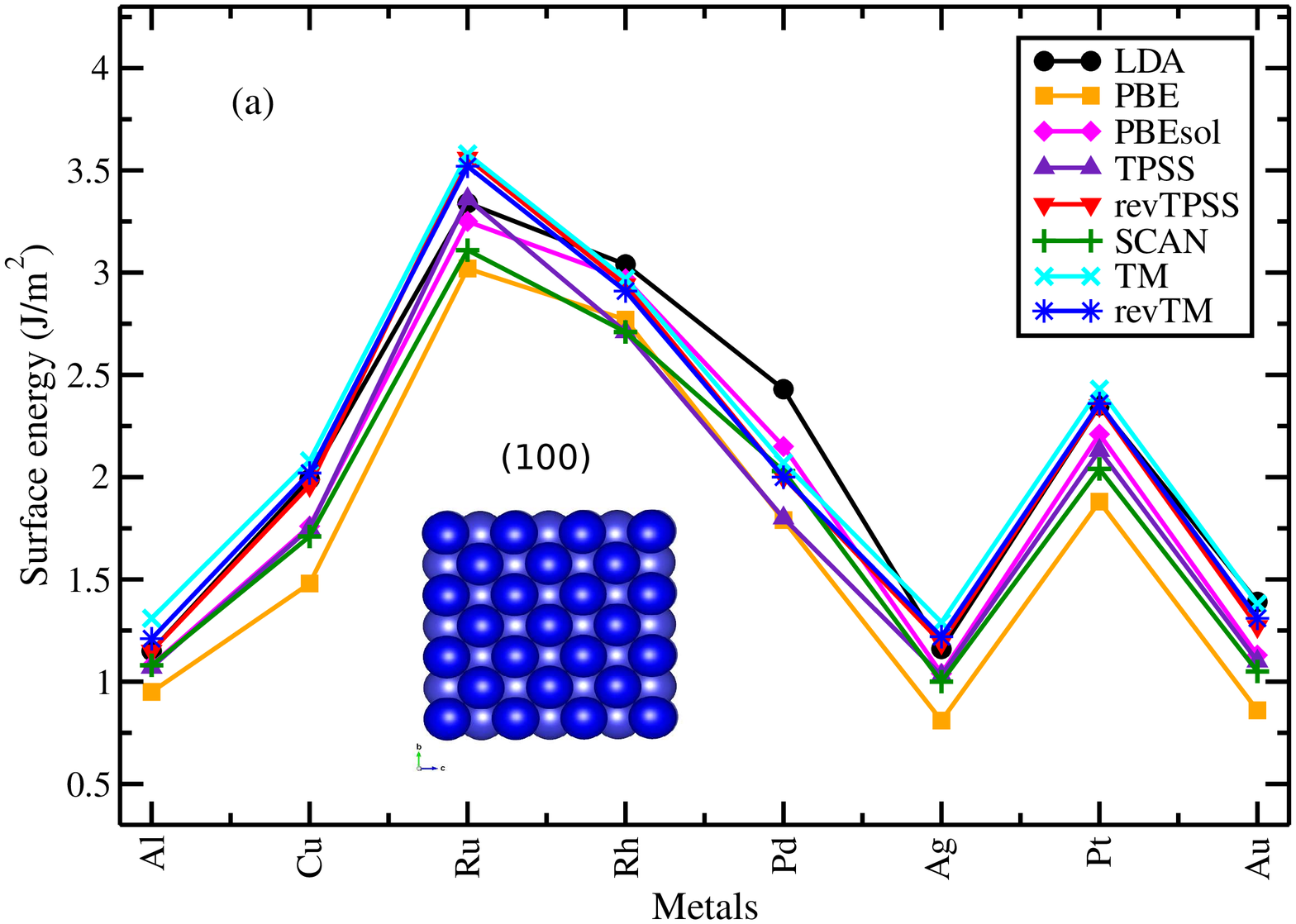}
\includegraphics[width=2.2in,height=2.2in,angle=0.0]{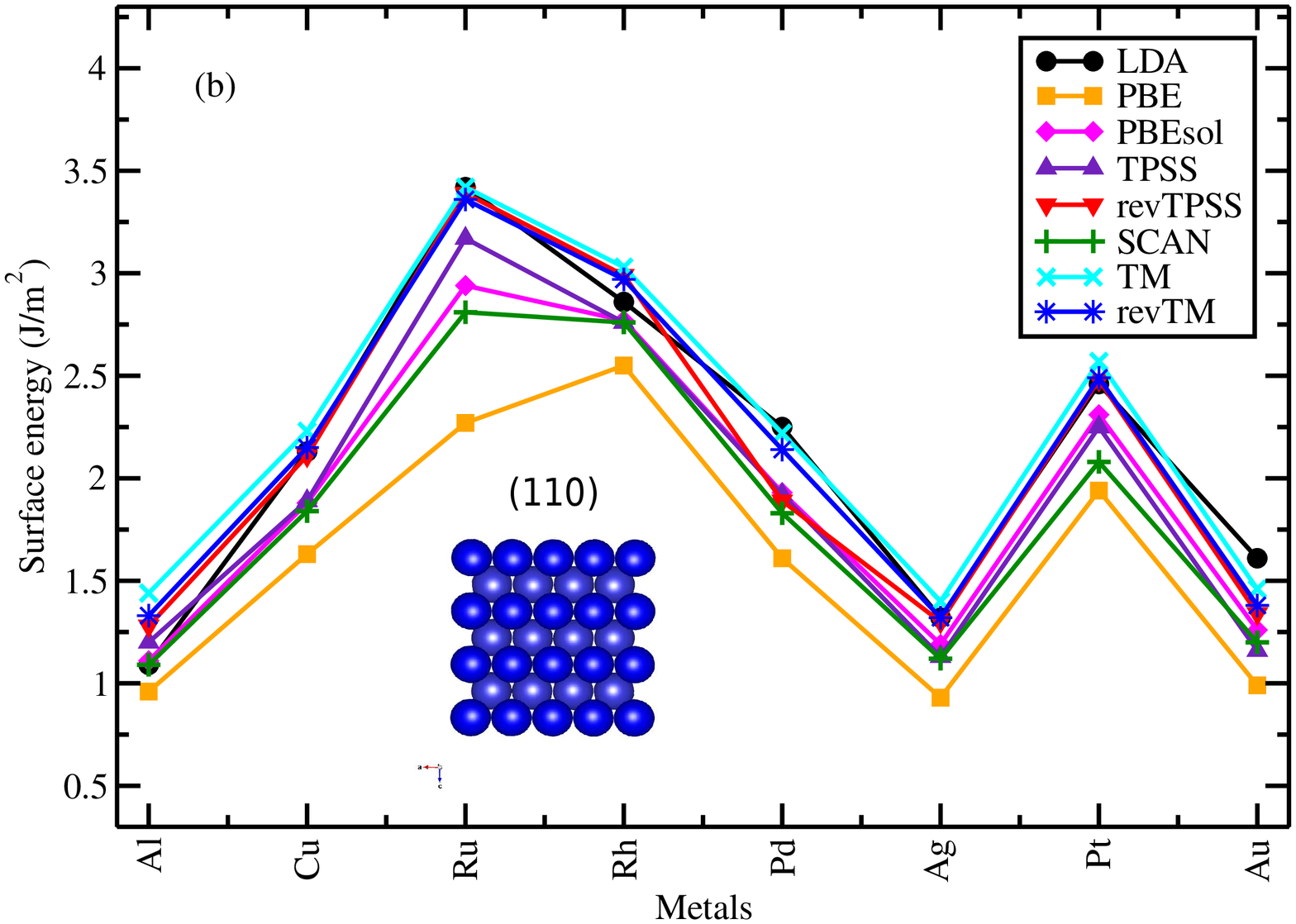}
\includegraphics[width=2.2in,height=2.2in,angle=0.0]{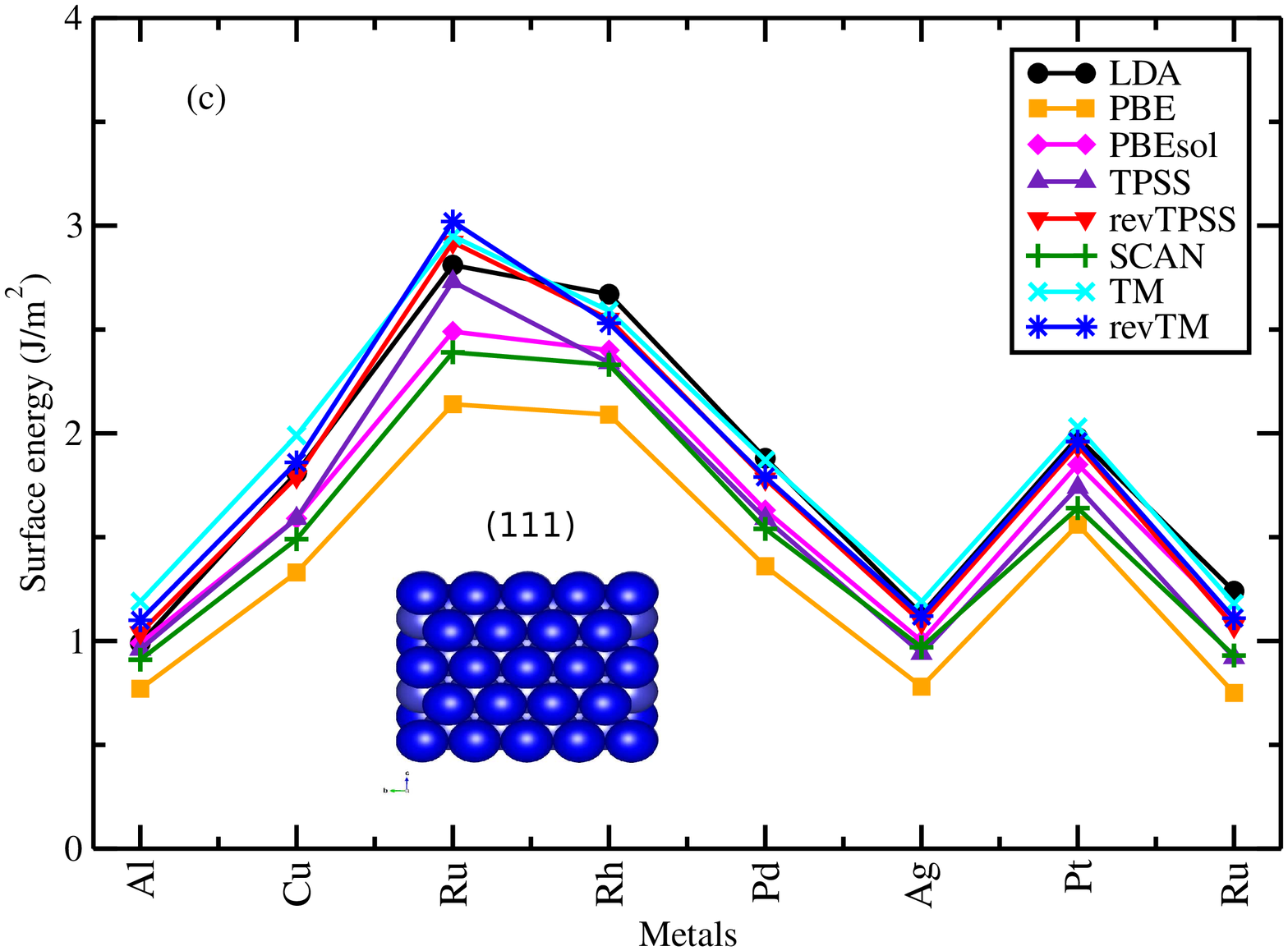}
\end{center}
\caption{Shown are the surface energies (J/m$^2$) for the selected metals using different functionals.}
\label{sur_eng2}
\end{figure*}

\begingroup
\squeezetable
\begin{table*}
\caption{\label{sur_eng}Metallic surface energy (J/m$^2$) computed using the considered functionals. The LDA, PBE, PBEsol, 
and SCAN functionals results and reference values are taken from ref.~\cite{PatraE9188}. The best values are in bold and 
the most deviating values are underlined.}
{\scriptsize
\begin{ruledtabular}
\begin{tabular}{cccccccccccccccccccccccccccccccccc}
\multicolumn{1}{l}{Metals} &
\multicolumn{1}{l}{Surfaces} &
\multicolumn{1}{c}{LDA} &
\multicolumn{1}{c}{PBE} &
\multicolumn{1}{c}{PBEsol} &
\multicolumn{1}{c}{TPSS} &
\multicolumn{1}{c}{revTPSS} &
\multicolumn{1}{c}{SCAN} &
\multicolumn{1}{c}{TM} &
\multicolumn{1}{c}{revTM} &
\multicolumn{1}{c}{Expt.}\\
\hline

	&	100	&	1.15	&	0.95	&	1.08	&	1.07	&	1.15	&	1.08	&	1.31	&	1.21	&		\\	
Al	&	110	&	1.09	&	0.96	&	1.11	&	1.20	&	1.28	&	1.09	&	1.44	&	1.33	&		\\	
	&	111	&	0.99	&	0.77	&	0.99	&	0.96	&	1.04	&	0.91	&	1.19	&	1.10	&		\\	[0.2cm]
	&	$\bar{\sigma}$	&	1.08	&	\underline{0.89}	&	1.06	&	1.08	&	\textbf{1.16}	&	1.03	&	1.31	&	1.21	&	1.14	\\	[0.4cm]
	&	100	&	1.99	&	1.48	&	1.76	&	1.75	&	1.96	&	1.71	&	2.08	&	2.02	&		\\	
Cu	&	110	&	2.13	&	1.63	&	1.88	&	1.89	&	2.11	&	1.84	&	2.23	&	2.15	&		\\	
	&	111	&	1.81	&	1.33	&	1.59	&	1.59	&	1.79	&	1.49	&	1.99	&	1.86	&		\\	[0.2cm]
	&	$\bar{\sigma}$	&	1.98	&	\underline{1.48}	&	\textbf{1.74}	&	\textbf{1.74}	&	1.95	&	1.68	&	\underline{2.10}	&	2.01	&	1.79	\\	[0.4cm]
	&	100	&	3.34	&	3.02	&	3.25	&	3.36	&	3.56	&	3.11	&	3.58	&	3.52	&		\\	
Ru	&	110	&	3.42	&	2.27	&	2.94	&	3.17	&	3.39	&	2.81	&	3.42	&	3.36	&		\\	
	&	111	&	2.81	&	2.14	&	2.49	&	2.73	&	2.92	&	2.39	&	2.95	&	3.02	&		\\	[0.2cm]
	&	$\bar{\sigma}$	&	3.19	&	\underline{2.48}	&	2.89	&	\textbf{3.09}	&	3.29	&	2.77 & 	3.32	&	3.30	&	3.04	\\	[0.4cm]
	&	100	&	3.04	&	2.77	&	2.97	&	2.71	&	2.94	&	2.71 & 	2.97	&	2.91	&		\\	
Rh	&	110	&	2.86	&	2.55	&	2.77	&	2.76	&	2.99	&	2.76& 	3.03	&	2.97	&		\\	
	&	111	&	2.67	&	2.09	&	2.40	&	2.34	&	2.55	&	2.33 & 2.59	&	2.53	&		\\	[0.2cm]
	&	$\bar{\sigma}$	&	\underline{2.86}	&	2.47	&	\textbf{2.71}	&	2.60	&	2.83	&	2.60	&	2.86	&	2.80	&	2.66	\\	[0.4cm]
	&	100	&	2.43	&	1.79	&	2.15	&	1.80	&	2.00	&	2.03	&	2.07	&	2.00	&		\\	
Pd	&	110	&	2.25	&	1.61	&	1.93	&	1.92	&	1.89	&	1.83	&	2.22	&	2.14	&		\\	
	&	111	&	1.88	&	1.36	&	1.63	&	1.59	&	1.78	&	1.54	&	1.87	&	1.79	&		\\	[0.2cm]
	&	$\bar{\sigma}$	&	2.19	&	\underline{1.59}	&	1.90	&	1.77	&	1.97	&	1.80    &	2.05	&	\textbf{1.98}	&	2.00	\\	[0.4cm]
	&	100	&	1.16	&	0.81	&	1.04	&	1.03	&	1.19	&	1.00	&	1.29	&	1.22	&		\\	
Ag	&	110	&	1.32	&	0.93	&	1.19	&	1.13	&	1.30	&	1.12	&	1.40	&	1.32	&		\\	
	&	111	&	1.13	&	0.78	&	1.00	&	0.94	&	1.09	&	0.97	&	1.19	&	1.12	&		\\	[0.2cm]
	&	$\bar{\sigma}$	&	1.20	&	\underline{0.84}	&	1.08	&	1.03	&	1.19	&	1.03	&	1.29	&	\textbf{1.22}	&	1.25	\\	[0.4cm]
	&	100	&	2.35	&	1.88	&	2.21	&	2.13	&	2.35	&	2.04	&	2.43	&	2.36	&		\\	
Pt	&	110	&	2.46	&	1.94	&	2.31	&	2.25	&	2.48	&	2.08	&	2.57	&	2.49	&		\\	
	&	111	&	1.98	&	1.56	&	1.85	&	1.74	&	1.94	&	1.64	&	2.03	&	1.96	&		\\	[0.2cm]
	&	$\bar{\sigma}$	&	2.26	&	\underline{1.79}	&	2.12	&	2.04	&	2.26	&	1.92	&	\textbf{2.34}	&	2.27	&	2.49	\\	[0.4cm]
	&	100	&	1.39	&	0.86	&	1.13	&	1.10	&	1.27	&	1.05	&	1.38	&	1.31	&		\\	
Au	&	110	&	1.61	&	0.99	&	1.26	&	1.16	&	1.34	&	1.20	&	1.46	&	1.38	&		\\	
	&	111	&	1.24	&	0.75	&	1.10	&	0.92	&	1.07	&   0.93	&	1.18	&	1.11	&		\\	[0.2cm]
	&	$\bar{\sigma}$	&	\textbf{1.41}	&	\underline{0.87}	&	1.16	&	1.06	&	1.23	&	1.06	&	1.34	&	1.27	&	1.51	\\	[0.2cm]

	\hline\hline
ME (J/m$^2$)	&		&	0.04	&	\underline{-0.43}	&	-0.15	&	-0.18	&	\textbf{0.00}	&	-0.25	&	0.09	&	0.02	&		\\
MAE (J/m$^2$)	&		&	\textbf{0.15}	&	\underline{0.43}	&	0.17	&	0.20	&	\textbf{0.15}	&	0.25	&	0.17	&	\textbf{0.15}	&		\\
MARPE (\%)	&		&	\textbf{7.21}	&	\underline{23.58}	&	9.16	&	11.12	&	7.42	&	13.40	&	8.99	&	7.55	&		\\

	\end{tabular}
\end{ruledtabular}
}
\end{table*}
\endgroup

\textbf{Bulk moduli :}~The accuracy of bulk moduli depends on the accuracy of the lattice constants. The performance 
of functionals for the bulk moduli are presented in Table~\ref{tab_bm} and MRPE is plotted in Fig.~\ref{fig5}. Here, 
we use third order Birch-Murnaghan equation of state to fit the energy-volume curve. Regarding the overall performances, 
the TM functional gives the least MAE followed by the revTM and SCAN functionals. The revTM improves the performance of 
TM functional for simple metals and ionic solids. This is because the lattice constants of the revTM functionals are 
obtained to be better for those solids. For the very same reason, the semiconductor bulk moduli are underestimated by 
revTM compared to the TM functional. Concerning the performance of other functionals, the tendency in the lattice constants 
is followed in the performance of bulk moduli. Except the SCAN and TM based functionals, other GGA and meta-GGA functionals 
show underestimation in results for most of the solids. Most pronounced underestimation is observed for PBE functional, 
while the LDA, as usual, overestimates the bulk moduli and shows tendency opposite to that of PBE. 

\begin{figure}
\begin{center}
\includegraphics[width=\columnwidth,height=7.0cm]{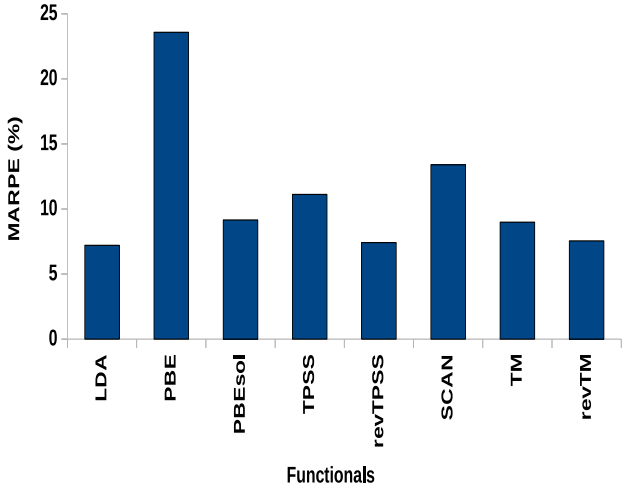}
\end{center}
\caption{Shown are the total MARPE of different functionals calculated using mean surface energies for selected metals.}
\label{marpe_sur}
\end{figure}

\textbf{Cohesive enrgies:}~Next, we assess the performance of revTM along with other functionals for the cohesive energies 
of solids. The performance of functionals are presented in Table~\ref{tab_coh} and plotted in Fig.~\ref{fig6}. Here, also 
we observe interesting performance for revTM compared to TM functionals as the revTM improves the performance over TM for 
most of the solids. Individual consideration shows that the revTM improves the performance for simple metals, transition 
metals, insulators, and semiconductors. For ionic solids, the performance of TM is slightly better than revTM. This is quite
natural because the change in correlation energy in revTM actually suits more than TM correlation and cancels error for 
second-order gradient approximation with correlation for the low-density limit. We observe that for most of the solids,
the revTM reduces the cohesive energies of the TM. Therefore, improves the overall performance of the cohesive energies. 
However, within meta-GGA functionals, the least MAE is observed using the TPSS functional. Also, the SCAN functional is 
quite good in predicting the cohesive energies compared to TM and revTM. The SCAN functional suits better for the 
simple metals, transition metals, and insulators in comparison to the revTM. Regarding the GGA based functionals. As usual, 
the PBE performs better compared to the PBEsol. This is due to the improved atomization energies of PBE compared to the PBEsol.  

\section{Real metallic surface energies}
Investigating any new XC functional performance for the real metallic surfaces are particularly interesting because 
those analysis has industrial applications. Several meta-GGA are developed along the Jacob ladder with increasing
accuracy in metal surface energy. Recent progress in this direction shows that the non-local functional SCAN+rVV$10$
\cite{PatraE9188} (SCAN functional plus revised Vydrov-van Voorhis non-local correlation functional) and quasi $2$D-GGA
functionals~\cite{PhysRevLett.108.126402} predict the most accurate metallic surface energies. However, in this work we 
assess the metallic surface energies of the semilocal XC functionals without including any non-local correlation. The
main aim of this paper is to assess the accuracy of LDA, PBE, PBEsol TPSS, revTPSS, SCAN, TM, and revTM functional for 
the metal surface energies. However, in our comparison we do not include the SCAN+rVV$10$ results and to the best of our
knowledge, the comparison of SCAN and TM functionals for the real metallic surfaces are done for the first time in this paper. 

But before going into the real application of the metallic surface energies, we consider the surface exchange-correlation 
energy of the jellium model~\cite{PhysRevB.84.045126}. The jellium serves as a model for the metallic systems. It consists 
of a homogeneous electron gas, where the electron charges are compensated with the positive background charge. The bulk
density of the jellium model remains constant and varies rapidly on the surface. Because in solids the valence electron 
density is slowly varying, the analysis of jellium surface XC energies helps to understand the accuracy of a particular 
XC energy for the solid-state systems. The underlying jellium surface exchange-correlation energy is defined as
\cite{PhysRevB.84.045126},
\begin{equation}
    \sigma_{xc}=\int_{-\infty}^{\infty}~dz~\rho(z)[\epsilon_{xc}[\rho;z]-\epsilon_{xc}^{unif}[\bar{\rho}]]~.
\end{equation}
In Table~\ref{tab5}, we present the jellium surface XC energies of the functionals under consideration. In this case, 
the diffusion Monte Carlo (DMC) results are considered as the reference values. An interesting observation is that the 
revTM reduces the jellium surface XC energies of the TM functional and predicts the values close to the DMC values. 
However, for these simple jellium systems, the LDA, GGA and other meta-GGA functionals are also quite accurate. The 
improvement in the jellium surface energies is well-known fact from the change in correlation as it is already mentioned 
in revTPSS paper~\cite{PhysRevLett.103.026403}. In their work, it is mentioned that the reduction of $\beta$ with $r_s$ 
actually decreases surface energies which in fact reflects in the performance of the revTM functional. Therefore, the 
change in the correlation is important and well suited with the TM functional.     

As a real test and assessment of the revTM functional, we consider the surface energies of the real metals compiled in 
ref~\cite{PatraE9188}. It is defined as the energy required to construct a surface from an infinite crystal. Computationally, 
it is measured by the formula,
\begin{equation}
    \sigma=\frac{1}{2A}\Big(E_N^{slab}-N\epsilon^{bulk}_{atom}\Big)~,
\end{equation}
where $E_N^{slab}$ is the total energy of the relaxed surface slab with $N$ atoms, and $\epsilon^{bulk}_{atom}$ is energy 
of the bulk with one atom and $A$ is the slab surface area.   

In practice, it is suggested to measure the mean of all surfaces to compare with the experimental surface energies. 
Therefore, we compare the surface energies taking the mean of the contribution of the values obtained from $(100)$, 
$(110)$, and $(111)$ surfaces. In Table~\ref{sur_eng}, we summarize the contribution of each surface energies. Here, 
the LDA, PBE, PBEsol, and SCAN values are taken from ref.~\cite{PatraE9188}. The TPSS, revTPSS, TM, and revTM surface 
energies are calculated in this work. In Fig.~\ref{sur_eng2} and Fig.~\ref{marpe_sur}, we plot the surface energy
contribution from each surface and MARPE respectively. Investigation of the contribution of the surface energies 
obtained from different surfaces using semilocal functionals, we remark that LDA gives quality surface energies due to
the well-known cancellation of error in exchange and correlation. The PBE underestimates the surface energies due to 
the inability of the recovering the exact second-order gradient approximation which indeed necessary for the good
surface energies. By recovering the exact second-order gradient approximation, the PBEsol improves the performance 
over PBE. While improving the correlation and more conveniently obeying the fourth-order gradient approximations 
over a broader range of density gradient approximations, the revTPSS improves the surface energies over TPSS. For 
the very same reason (as applied in the correlation) we observe the improvement in the surface energies for revTM 
over TM functional. Overall consideration shows that the LDA, revTPSS, and revTM functional agree very well with 
the experimental values. Interesting observation is that the TM functional overestimates the surface energies for 
most of the metals which are actually corrected by its revTM counterpart. 

\section{Conclusions} 
In this work, we proposed a revised TM (revTM) functional by improving the slowly varying density correction of the 
exchange and incorporating the revTPSS like correlation energy in the functional form. The proposed modification over 
TM functional is assessed for the solid-state lattice constants, bulk moduli, cohesive energies, jellium surface 
exchange-correlation energies and surface energies of the metals. Performance of the revTM functional shows its 
accuracy over broad range of the molecular and solid-state systems. Specifically, it improves the cohesive energies, 
jellium surface energy-correlation energies, surface energies of real metals and molecular atomization energies by
maintaining the overall accuracy of the lattice constants and bulk moduli. As the physical insight of the revTM 
functional performances, one can say that both the molecular and solid-state systems are treated in a more balanced 
way because it works well both for the localized (e.g., AE6 atomization energies) and delocalized systems (solid-state
physics). It's because the density matrix expansion based exchange hole is localized by nature and proper incorporation 
of the slowly varying fourth-order gradient expansion. The present modification keeps all the good properties of the 
TM functional, additionally, it improves the functional performances in the low-density limit through a simple
revision. It is also noticed that the present modifications treat both the quantum chemical and solid-state properties 
in a more balanaced way than other accurate and widely appreciated meta-GGA like SCAN functional.     

As a concluding remark, we can say that the revTPSS and revTM have one thing in common i.e. both cancel exchange and 
correlation at the low-density limit. However, the construction of the TM exchange functional is different from those
constructed only from the fourth-order gradient approximation. In the case of TM exchange functional, part of the 
exchange hole is constructed from density matrix expansion which is more localized than the convenient exchange hole 
obtained from the fourth-order gradient approximation. Therefore, it is desirable to cancel the error in the low-density 
($r_s\to \infty$) limit when the more localized density matrix expansion based exchange hole is coupled with the exchange 
hole of the fourth-order gradient approximation which indeed holds in the revTM functional. However, more conventionally 
and quite physically motivated $\beta$ form can also be derived for the TM exchange functional which is a matter of our 
future study. 

\section{Acknowledgement}
S.J. and P.S. would like to acknowledge and thank Dr. Lucian A. Constantin for providing some computational resources 
and useful comments regarding the work. S.J. would also like to acknowledge the financial support from the Department 
of Atomic Energy, The government of India. K.S. would like to acknowledge the financial support from the Department 
of Science and Technology, Government of India, during his summer internship in NISER under the supervision of P.S..

\bibliography{citations}

\end{document}